\RequirePackage[2020-02-02]{latexrelease}
\documentclass[twocolumn, times]{aastex631}
\usepackage{mathptmx}
\pdfoutput=1

\newcommand{\fCAL}{10$^{-16}$\,erg\,s$^{-1}$\,cm$^{-2}$}

\newcommand{\stmetL}{\ifmmode [{\rm Z}/{\rm H}]_{\rm L} \else [Z/H]$_{\rm L}$\fi\xspace}
\newcommand{\stmetM}{\ifmmode [{\rm Z}/{\rm H}]_{\rm M} \else [Z/H]$_{\rm M}$\fi\xspace}
\newcommand{\stageL}{\ifmmode \left<\log \age\right>_{\rm L} \else $\left<\log \age\right>_{\rm L}$\fi\xspace}
\newcommand{\stageM}{\ifmmode \left<\log \age\right>_{\rm M} \else $\left<\log \age\right>_{\rm M}$\fi\xspace}

\DeclareRobustCommand{\ion}[2]{%
\relax\ifmmode
\ifx\testbx\f@series
{\mathbf{#1\mathsc{#2}}}\else
{\mathrm{#1\mathsc{#2}}}\fi
\else\textup{#1{\mdseries\textsc{#2}}}%
\fi}

\newcommand{\Hii}{\ion{H}{II}}
\newcommand{\HII}{\Hii}
\newcommand{\Halpha}{\ion{H}{$\alpha$}}
\newcommand{\oiii}{[\ion{O}{iii}]}
\newcommand{\oii}{[\ion{O}{ii}]}
\newcommand{\siii}{[\ion{S}{iii}]}
\newcommand{\sii}{[\ion{S}{ii}]}
\newcommand{\nii}{[\ion{N}{ii}]}

\submitjournal{AJ}

\begin{document}

\author[0000-0002-7339-3170]{Niv Drory}
\affiliation{McDonald Observatory, The University of Texas at Austin, 1 University Station, Austin, TX 78712-0259, USA}
\email{drory@astro.as.utexas.edu}
\correspondingauthor{Niv Drory}

\author[0000-0003-4218-3944]{Guillermo A.\ Blanc}
\affiliation{Observatories of the Carnegie Institution for Science, 813 Santa Barbara Street, Pasadena, CA 91101, USA}
\affiliation{Departamento de Astronom\'{i}a, Universidad de Chile, Camino del Observatorio 1515, Las Condes, Santiago, Chile}

\author[0000-0001-6551-3091]{Kathryn Kreckel}
\affiliation{Astronomisches Rechen-Institut, Zentrum f\"ur Astronomie der Universit\"at Heidelberg, M\"onchhofstr.\ 12-14, D-69120 Heidelberg, Germany}

\author[0000-0001-6444-9307]{Sebasti\'an F.\ S\'anchez}
\affiliation{Instituto de Astronom\'{\i}a, Universidad Nacional Aut\'onoma de M\'exico, A.P.\ 106, Ensenada 22800, BC, M\'exico}

\author[0000-0001-6444-9307]{Alfredo Mej\'{\i}a-Narv\'aez}
\affiliation{Departamento de Astronom\'{i}a, Universidad de Chile, Camino del Observatorio 1515, Las Condes, Santiago, Chile}

\author[0000-0002-2368-6469]{Evelyn J.\ Johnston}
\affiliation{Instituto de Estudios Astrof\'isicos, Facultad de Ingenier\'ia y Ciencias, Universidad Diego Portales, Av.\ Ej\'ercito Libertador 441, Santiago, Chile}

\author[0000-0002-2262-8240]{Amy M.\ Jones}
\affiliation{Space Telescope Science Institute, 3700 San Martin Drive, Baltimore, MD 21218, USA}

\author{Eric W.\ Pellegrini}
\affiliation{Universit\"{a}t Heidelberg, Zentrum f\"{u}r Astronomie, Institut f\"{u}r Theoretische Astrophysik, Albert-Ueberle-Str.\ 2, D-69120 Heidelberg, Germany}

\author[0000-0003-1905-2815]{Nicholas P.\ Konidaris}
\affiliation{Observatories of the Carnegie Institution for Science, 813 Santa Barbara Street, Pasadena, CA 91101, USA}

\author{Tom Herbst}
\affiliation{Max Planck Institute for Astronomy, K\"onigstuhl 17, D-69117 Heidelberg, Germany}

\author[0000-0003-2486-3858]{Jos\'e S\'anchez-Gallego}
\affiliation{Department of Astronomy, University of Washington, Box 351580, Seattle, WA 98195, USA}

\author[0000-0001-9852-1610]{Juna A.\ Kollmeier}
\affiliation{Canadian Institute for Theoretical Astrophysics, University of Toronto, Toronto, ON M5S-98H, Canada}
\affiliation{Observatories of the Carnegie Institution for Science, 813 Santa Barbara Street, Pasadena, CA 91101, USA}

%
%
\author{Florence de Almeida}
\affiliation{Instituto de Estudios Astrof\'isicos, Facultad de Ingenier\'ia y Ciencias, Universidad Diego Portales, Av.\ Ej\'ercito Libertador 441, Santiago, Chile}

\author{Jorge K.\ Barrera-Ballesteros}
\affiliation{Instituto de Astronom\'{\i}a, Universidad Nacional Aut\'onoma de M\'exico, A.P.\ 70-264, 04510 M\'exico D.\ F., M\'exico}

\author[0000-0002-3601-133X]{Dmitry Bizyaev}
\affiliation{Apache Point Observatory and New Mexico State University, P.O.\ Box 59,
Sunspot, NM 88349-0059, USA}

\author[0000-0002-8725-1069]{Joel R.\ Brownstein}
\affiliation{Department of Physics and Astronomy, University of Utah, 115 S.\ 1400 E., Salt Lake City, UT 84112, USA}

\author{Mar Canal i Saguer}
\affiliation{Astronomisches Rechen-Institut, Zentrum f\"ur Astronomie der Universit\"at Heidelberg, M\"onchhofstr.\ 12-14, D-69120 Heidelberg, Germany}

\author[0000-0002-4289-7923]{Brian Cherinka}
\affiliation{Space Telescope Science Institute, 3700 San Martin Drive, Baltimore, MD 21218, USA}

\author[0000-0002-6797-696X]{Maria-Rosa L.\ Cioni}
\affiliation{Leibniz-Institut f\"ur Astrophysik Potsdam, An der Sternwarte 16, D-14482 Potsdam, Germany}

\author[0000-0002-8549-4083]{Enrico Congiu}
\affiliation{European Southern Observatory, Alonso de C\'ordova 3107, Casilla 19, Santiago 19001, Chile}

\author[0000-0002-2248-6107]{Maren Cosens}
\affiliation{Observatories of the Carnegie Institution for Science, 813 Santa Barbara Street, Pasadena, CA 91101, USA}

\author[0000-0003-4254-7111]{Bruno Dias}
\affiliation{Universidad Andres Bello, Departamento de Ciencias Físicas, Instituto de Astrofísica, Av. Fernández Concha 700, Santiago, Chile}

\author{John Donor}
\affiliation{Department of Physics \& Astronomy, Texas Christian University, Fort Worth, TX 76129, USA}

\author[0000-0002-4755-118X]{Oleg Egorov}
\affiliation{Astronomisches Rechen-Institut, Zentrum f\"ur Astronomie der Universit\"at Heidelberg, M\"onchhofstr.\ 12-14, D-69120 Heidelberg, Germany}

\author[0000-0003-2717-8784]{Evgeniia Egorova}
\affiliation{Astronomisches Rechen-Institut, Zentrum f\"ur Astronomie der Universit\"at Heidelberg, M\"onchhofstr.\ 12-14, D-69120 Heidelberg, Germany}

\author[0000-0001-8499-2892]{Cynthia S.\ Froning}
\affiliation{Southwest Research Institute, 6220 Culebra Road, San Antonio, TX 78238  }

\author[0000-0002-8586-6721]{Pablo Garc\'ia}
\affiliation{Instituto de Astronomía, Universidad Católica del Norte, Av.\ Angamos 0610, Antofagasta, Chile}
\affiliation{Chinese Academy of Sciences South America Center for Astronomy, National Astronomical Observatories, CAS, Beijing 100101, China}

\author[0000-0001-6708-1317]{Simon C. O. Glover}
\affiliation{Universit\"{a}t Heidelberg, Zentrum f\"{u}r Astronomie, Institut f\"{u}r Theoretische Astrophysik, Albert-Ueberle-Str.\ 2, D-69120 Heidelberg, Germany}

\author[0009-0001-4629-8098]{Hannah Greve}
\affiliation{Astronomisches Rechen-Institut, Zentrum f\"ur Astronomie der Universit\"at Heidelberg, M\"onchhofstr.\ 12-14, D-69120 Heidelberg, Germany}

\author[0000-0002-5844-4443]{Maximilian H\"aberle}
\affiliation{Max Planck Institute for Astronomy, K\"onigstuhl 17, D-69117 Heidelberg, Germany}

\author{Kevin Hoy}
\affiliation{Instituto de Estudios Astrof\'isicos, Facultad de Ingenier\'ia y Ciencias, Universidad Diego Portales, Av.\ Ej\'ercito Libertador 441, Santiago, Chile}
\affiliation{Millennium Nucleus on Young Exoplanets and their Moons (YEMS)}

\author[0000-0002-9790-6313]{Hector Ibarra}
\affiliation{Instituto de Astronom\'{\i}a, Universidad Nacional Aut\'onoma de M\'exico, A.P.\ 70-264, 04510 M\'exico D.\ F., M\'exico}

\author[0000-0002-4825-9367]{Jing Li}
\affiliation{Astronomisches Rechen-Institut, Zentrum f\"ur Astronomie der Universit\"at Heidelberg, M\"onchhofstr.\ 12-14, D-69120 Heidelberg, Germany}

\author[0000-0002-0560-3172]{Ralf S.\ Klessen}
\affiliation{Universit\"{a}t Heidelberg, Zentrum f\"{u}r Astronomie, Institut f\"{u}r Theoretische Astrophysik, Albert-Ueberle-Str.\ 2, D-69120 Heidelberg, Germany}
\affiliation{Universit\"{a}t Heidelberg, Interdisziplin\"{a}res Zentrum f\"{u}r Wissenschaftliches Rechnen, Im Neuenheimer Feld 205, D-69120 Heidelberg, Germany}

\author[0000-0002-7955-7359]{Dhanesh Krishnarao}
\affiliation{Department of Physics, Colorado College, 14 East Cache la Poudre St., Colorado Springs, CO, 80903, USA}

\author[0000-0002-5320-2568]{Nimisha Kumari}
\affiliation{Space Telescope Science Institute, 3700 San Martin Drive, Baltimore, MD 21218, USA}

\author[0000-0002-4134-864X]{Knox S.\ Long}
\affiliation{Space Telescope Science Institute, 3700 San Martin Drive, Baltimore, MD 21218, USA}

\author[0000-0002-6972-6411]{Jos\'e Eduardo M\'endez-Delgado}
\affiliation{Astronomisches Rechen-Institut, Zentrum f\"ur Astronomie der Universit\"at Heidelberg, M\"onchhofstr.\ 12-14, D-69120 Heidelberg, Germany}

\author[0009-0006-3186-5826]{Silvia Anastasia Popa}
\affiliation{Astronomisches Rechen-Institut, Zentrum f\"ur Astronomie der Universit\"at Heidelberg, M\"onchhofstr.\ 12-14, D-69120 Heidelberg, Germany}

\author{Solange Ramirez}
\affiliation{Observatories of the Carnegie Institution for Science, 813 Santa Barbara Street, Pasadena, CA 91101, USA}

\author[0000-0003-4996-9069]{Hans-Walter Rix}
\affiliation{Max Planck Institute for Astronomy, K\"onigstuhl 17, D-69117 Heidelberg, Germany}

\author{Aurora Mata S\'anchez}
\affiliation{Instituto de Astronom\'{\i}a, Universidad Nacional Aut\'onoma de M\'exico, A.P.\ 70-264, 04510 M\'exico D.\ F., M\'exico}

\author[0000-0001-8858-1943]{Ravi Sankrit}
\affiliation{Space Telescope Science Institute, 3700 San Martin Drive, Baltimore, MD 21218, USA}

\author[0000-0002-8883-6018]{Natascha Sattler}
\affiliation{Astronomisches Rechen-Institut, Zentrum f\"ur Astronomie der Universit\"at Heidelberg, M\"onchhofstr.\ 12-14, D-69120 Heidelberg, Germany}

\author[0000-0002-4454-1920]{Conor Sayres}
\affiliation{Department of Astronomy, University of Washington, Box 351580, Seattle, WA 98195, USA}

\author{Amrita Singh}
\affiliation{Universidad de Chile, Av.\ Libertador Bernardo O'Higgins 1058, Santiago de Chile, Chile}

\author[0000-0003-1479-3059]{Guy Stringfellow}
\affiliation{Department of Astrophysical and Planetary Sciences, University of Colorado, 389 UCB, Boulder, CO 80309-0389, USA}

\author{Stefanie Wachter}
\affiliation{Observatories of the Carnegie Institution for Science, 813 Santa Barbara Street, Pasadena, CA 91101, USA}

\author[0000-0002-7365-5791]{Elizabeth Jayne Watkins}
\affiliation{Jodrell Bank Centre for Astrophysics, Department of Physics and Astronomy, University of Manchester, Oxford Road, Manchester M13 9PL, UK}

\author[0000-0002-0786-7307]{Tony Wong}
\affiliation{Department of Astronomy, University of Illinois, Urbana, IL 61801, USA}

\author[0000-0001-8289-3428]{Aida Wofford}
\affiliation{Instituto de Astronom\'ia, Universidad Nacional Aut\'onoma de M\'exico, Unidad Acad\'emica en Ensenada, Km 103 Carr. Tijuana$-$Ensenada, Ensenada, B.C.,\\C.P. 22860, M\'exico}

\title{The SDSS-V Local Volume Mapper (LVM): Scientific Motivation and Project Overview}


\begin{abstract}
We present the Sloan Digital Sky Survey V (SDSS-V) Local Volume Mapper (LVM). The LVM is an integral-field spectroscopic survey of the Milky Way, Magellanic Clouds, and of a sample of local volume galaxies, connecting resolved pc-scale individual sources of feedback to kpc-scale ionized interstellar medium (ISM) properties. The 4-year survey covers the southern Milky Way disk at spatial resolutions of 0.05 to 1~pc, the Magellanic Clouds at 10~pc resolution, and nearby large galaxies at larger scales totaling $>4300$ sq.deg.\ of sky and more than 55M spectra. It utilizes a new facility of alt-alt mounted siderostats feeding 16~cm refractive telescopes, lenslet-coupled fiber-optics, and spectrographs covering 3600-9800\AA\ at $R\sim4000$. The ultra-wide field IFU has a diameter of 0.5~degrees with 1801 hexagonally packed fibers of 35.3~arcsec apertures. The siderostats allow for a completely stationary fiber system, avoiding instability of the line spread function seen in traditional fiber feeds. Scientifically, LVM resolves the regions where energy, momentum, and chemical elements are injected into the ISM at the scale of gas clouds, while simultaneously charting where energy is being dissipated (via cooling, shocks, turbulence, bulk flows, etc.) to global scales. This combined local and global view enables us to constrain physical processes regulating how stellar feedback operates and couples to galactic kinematics and disk-scale structures, such as the bar and spiral arms, as well as gas in- and out-flows.
\end{abstract}


\section{Introduction}\label{sec:intro}

The Local Volume Mapper (LVM) is one of three programs that comprise the fifth generation of the Sloan Digital Sky Survey\footnote{https://www.sdss.org} \citep[SDSS-V][]{kollmeier19,kollmeier24}. The goal of the LVM is to produce wide-area, high spatial resolution, integral-field spectroscopy (IFS) over the bulk of the southern Milky Way (MW) plane, a sparse grid of diffuse ionized gas (DIG) dominated high galactic latitude tiles, the Large Magellanic Cloud (LMC), and the Small Magellanic Cloud (SMC)\footnote{https://sdss.org/dr18/lvm/about/}. The survey also covers a sample of nearby galaxies. This is achieved by means of a small telescope (16\,cm diameter) equipped with a wide-field integral-field unit (IFU) consisting of 1801 microlens-coupled science fibers arranged in a hexagon of 0.5\,degrees diameter, and a 4-year observing program covering $>$4300 square degrees of the sky in $\sim$26000 survey tiles. The IFU is feeding 3 DESI-like spectrographs \citep{10.1117/12.2311996,desi22}, covering the wavelength range of 3600--9800\,\AA\ with a spectral resolution of $R\sim4000$ at \Halpha. 

A new observing facility called the Local Volume Mapper Instrument (LVM-I) consisting of four telescopes, fiber system, calibration system, and spectrographs has been built for this project at Las Campanas Observatory (LCO)\footnote{LCO is owned and operated by The Observatories of the Carnegie Institution for Science.}. A full description of the LVM-I is given by \citet{konidaris20}. 

The LVM-I wavelength range covers all optical nebular emission lines typically used as diagnostics of the physical conditions in the ionized interstellar medium (ISM), allowing us to map the thermal, chemical, photo-ionization and kinematic structure of ionized nebulae and the warm ($T\sim 10^3-10^4$\,K) ionized medium (WIM). Together with other datasets from the X-ray to the radio regime which in recent years have mapped the Milky Way disk with similar or greater coverage and on comparable spatial scales, the LVM dataset provides a global view of the Milky Way's ionized ISM while at the same time resolving the internal structure of individual star-forming regions (\HII\ regions) and star clusters.

In the last few decades, we have amassed a vast amount of observations revealing a number of correlations among galaxies' local and global properties \citep{Blanton2009,Sanchez2020}. These correlations open a window into the physics of galaxy formation that we must explain within the cosmological standard model.  While the role of dark matter is better understood in terms of fixing the history of galaxies' mass assembly and their first-order internal kinematics, the baryonic physics that shapes the coupled evolution of stars and gas still eludes us.  We know on kiloparsec scales that the formation of stars scales with the density of gas, and that the phase and motions of the gas are affected by the subsequent evolution of the stars \citep[e.g.][and references therein]{KE2012}.  However, these correlations remain merely descriptive in almost all extragalactic contexts.  Their existence is well-established, but their origins are uncertain.  Indeed, for explaining the emergence of these empirical correlations there are typically multiple and degenerate plausible physical scenarios \citep[e.g.][]{Kruijssen2019}.  

In contrast to past and present galaxy IFU surveys, e.g.\ CALIFA \citep{2012A&A...538A...8S}, SAMI \citep{sami}, MaNGA \citep{bundy15}, and PHANGS-MUSE \citep{Emsellem2022}, LVM prioritizes covering very few objects, the Galaxy and its nearest neighbors, but resolving them down to the physical scales from which the global correlations arise, to witness the direct interaction between the ISM and the sources that set its chemo-thermodynamics, while compromising on the number of galaxies.  Specifically, the LVM aims to map the ionized ISM on global scales of galactic outflows, fountains and winds, down to individual sources of feedback. LVM is resolving the scales on which energy, momentum, and chemical elements are injected into the ISM (mainly from high-mass stars on sub-pc to 10\,pc scales). It covers populations of gas clouds (50--100\,pc), and scales where energy is being dissipated (cooling, shocks, turbulence; 100\,pc -- 1\,kpc) and beyond to scales of bulk motions ($>$ 1\,kpc). The LVM's large survey area also allows observations of how these processes couple to the global scales of galactic kinematics and disk-scale structures such as the bar and spiral arms. Finally, LVM witnesses scales of galactic gas inflows and outflows (10\,kpc).

In this paper, we give a birds-eye overview of the project, starting with the scientific rationale followed by a high-level review of the instrument, survey, and the (public) data products. This paper serves as an introduction to the LVM for scientists wishing to engage with the data. This article is accompanied by a series of further technical publications giving in-depth descriptions of the LVM-I instrument \citep{konidaris20}; the telescope system \cite{Herbst22,Herbst2024}; survey design, data flow, and execution \citep{blanc24}; survey strategy, scheduling and operations \citep{johnston24}; data simulator \citep{Congiu24}; data reduction pipeline \citep[DRP,][]{mejia24}; spectro-photometric calibration \citep{kreckel24}; sky subtraction \citep{jones24}; and finally the data analysis pipeline \citep[DAP,][]{sanchez24}. 

In Section~\ref{sec:motivation}, we provide a review of LVM's science motivation and goals. This is followed by Section~\ref{sec:requirements}, where we describe the science requirements that lead to the definition of the instrument and survey. In Section~\ref{sec:instrument}, we briefly review the LVM Instrument (LVM-I), followed by an overview of our survey and calibration strategy as well as data products, reduction, and analysis (Section~\ref{sec:lvmsurvey}). We showcase the kind of data LVM provides using early survey data and describe example data products in Section~\ref{sec:data}. We end by summarizing the article in Section~\ref{sec:summary}.

\section{Scientific Motivation} \label{sec:motivation}

\subsection{Science Case Overview}
\begin{figure*}[ht]
\plotone{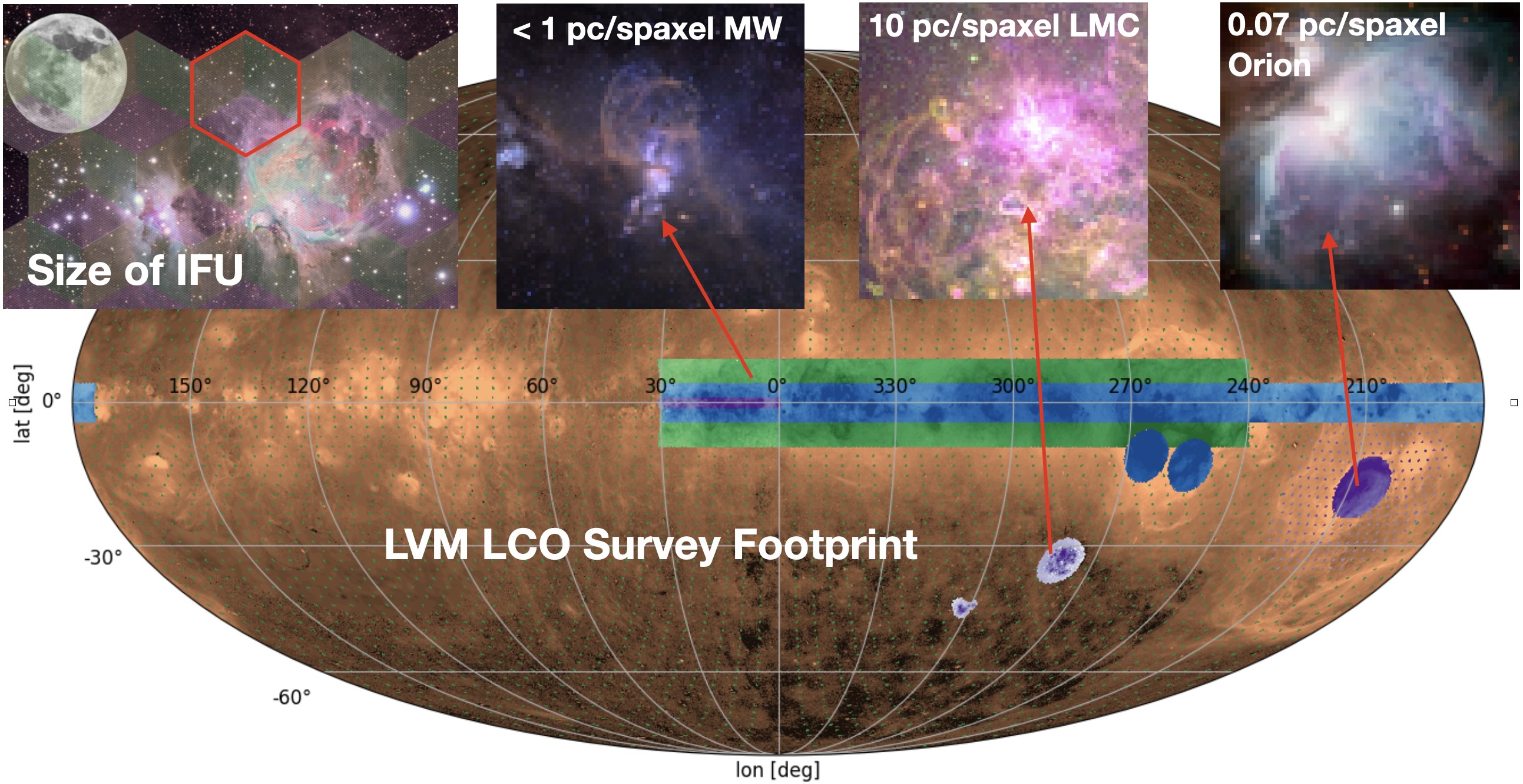}
\caption{LVM survey footprint overlaid on the \citet{fink03} \Halpha\ map of the full sky. We show different target classes in different colors for clarity. The Milky Way survey consists of an 8-degree band in the midplane of the disk spanning 30$\degr$ longitude through the Galactic center to 175$\degr$ (blue). A lower-priority 18-degree band spans 30$\degr$ longitude to 240$\degr$ (green). These bands around the midplane are augmented by coverage of the two nearby \HII\ regions: Orion (Purple) and Gum (Dark Blue). We supplement disk plane survey with a sparse grid (green) tracking large scale diffuse ionized gas above the Milky Way plane. LVM furthermore covers the LMC and SMC to their optical radius ($\sim R_{25}$; Light Purple). The simple geometric selection function of the MW midplane survey facilitates easy comparison with simulations. Zooming into the Orion region (lower right), we show ionized emission sampled at the LVM spaxel size of 0.07\,pc. At larger distances in the Milky Way of $\sim$3\,kpc, we still recover filamentary emission features (lower left) at 0.6\,pc resolution. Within the LMC (center bottom), we show a continuum \& ionized emission in the 30~Doradus star forming region, sampled at the 10\,pc spaxel size LVM will deliver in the Magellanic System.}\label{fig:overview}
\end{figure*}

The physical processes shaping galactic disks span orders of magnitude in physical scale. Star formation involves cooling warm atomic gas ($\gg1$\,kpc), assembling giant molecular clouds (GMCs; $\sim50$\,pc), and collapsing/fragmenting filaments, clumps, and cores ($\sim10-0.1$\,pc) inside GMCs, linking kpc to sub-pc scales \cite[e.g.][and references therein]{Schinnerer2019}. The effects of feedback by stellar winds, supernovae and supernova remnants, photo-ionization, radiation pressure, and AGN propagate from small to large scales, eventually giving rise to galaxy-wide phenomena such as chemical abundance gradients and azimuthal variations, galactic outflows, fountains, and the regulation of the physical conditions of the warm and diffuse ISM. Timescales span from the short free-fall and crossing times of clumps and cores, to the long dynamical times of galaxy disks \citep{Kennicutt1989,Elmegreen1993}. It is now clear that star formation must be at least partly a feedback-driven self-regulated process \citep[see, e.g.,][]{Ostriker2010,Crain2023}. Despite its importance, feedback is poorly understood \textbf{\textit{quantitatively}} \citep[see, e.g.,][]{Klessen2016}. How strongly, how far-reaching, and for how long does it affect its environment? How does this change with stars of different masses, ages and metal abundances? And how does it depend on the properties of the gas on which it acts (e.g.\ its clumpiness, ionization state, and kinematics)?

Most observational approaches adopt either the top-down view of 100\,pc to kpc scale maps of external galaxies \citep[e.g.][]{leroy2021, Emsellem2022}, or the bottom-up perspective of studying detailed parsec-scale structures of local regions in the MW and the Local Group \citep[e.g.][]{Kong2018, Pabst2020, Orozco-Duarte2023}. However, large and small scales are closely coupled. For example, turbulence is driven at the $~100-500$\,pc scale by various dynamical processes and cascades down to sub-parsec structures, while energy is injected at small $1-50$\,pc scales via stellar feedback that propagates to drive large scale bulk motions (even galactic outflows in the most extreme cases) that provide turbulent pressure support to the ISM.

The warm ($10^3-10^4$\,K) ionized phase of the ISM is an ideal tracer to study the effects of feedback at optical wavelengths \citep[see][for a review of the low density component]{Haffner2009}. Emission line diagnostics are at the core of measuring the physical condition of the ISM (e.g., densities, temperatures, dynamics, chemical abundances, dust extinction, ionization and excitation, etc.). On scales of $\approx 10 - 50$\,pc, relatively dense \Hii\ regions reprocess about half of the emitted ionizing stellar radiation, while the rest escapes and gets absorbed by diffuse ionized gas. On scales up to kiloparsecs, momentum, energy and metals get deposited, dissipated and distributed in the ISM.

LVM is designed to provide maps of the ISM resolving individual sources of feedback while at the same time covering whole galactic systems. The aim of the LVM is to map both the microscopic and macroscopic processes that regulate the behavior, structure, and longevity of the ionized ISM, starting with clouds in the MW and Local Group and their interface to the diffuse ISM to understanding the emergence of kpc-scale patterns. The overarching goal is to contribute to a synthesized multi-wavelength picture of the local and global physics of galactic disks, including molecular, atomic, and ionized gas, the stellar populations, SF, feedback and resulting gas flows, dust content, element abundances and enrichment, galactic environment, and local and global gravitational potential and kinematics. At the same time, LVM will provide important constraints to models of massive-star evolution and atmospheres. 

LVM provides spatially resolved spectroscopic observations of dense ionized nebulae, mainly \Hii\ regions but also planetary nebuale (PNe), supernova remnants (SNR), and superbubbles carved by the collective winds and SN explosions of massive stars, for reliable instantaneous measurements of their physical conditions, with full galactic coverage to obtain a population sample that enables us to understand the time evolution of similar regions as well as the variability of conditions among coeval regions. Separating the spatial and temporal dependencies in such a way for different physical processes involved in star formation and the injection of feedback is crucial to overcome the inherent degeneracies associated with observing instantaneous snapshots of a dynamic ISM at any given time. 

Clearly, optical emission lines trace only the warm ionized phase of the ISM. However, in recent years a number of large-area surveys of the Milky Way and Magellanic System have been conducted at many wavelengths and increasingly at spatial resolutions very similar to the LVM's. This rich set of data will be invaluable when addressing the physical state of the ISM. In the IR, the Milky Way has been imaged at 33\arcsec\ by BOLOCAM \citep{BGPS}, by the Spitzer Space Telescope \citep{GLIMPSE}, and by Herschel \citep{HiGAL}. Planck has provided maps of the dust distribution and polarization \citep{PlanckXIX}. 
Large portions of the galactic plane have been recently imaged in the sub-mm with a 19.2\arcsec\ beam by \citet{ATLASGAL}, in CO with 48\arcsec\ beam size by \citet{BUFCRAO}, in HI at 1\arcmin\ resolution by \citet{VGPS}, at 20\arcsec\ by \citet{THOR}, and at 2\arcmin\ by \citet{SGPS}. In the radio continuum, among others, recent surveys were carried out by the GMRT at 150\,MHz \citep{Intema2017} and  MeerKAT at 1.3\,GHz \citep{Goedhart2023}, as well as by the THOR project at the VLA \citep{THOR}.

Higher density gas at warm temperatures has been observed with a 2\arcsec\ beam through the stacking of several radio recombination lines by the Green Bank Telescope (GBT) in the GBT Diffuse Ionized Gas Survey \citep[GDIGS;][]{Luisi2020}. Additionally, the Wisconsin H-Alpha Mapper (WHAM) has an all-sky map of \Halpha\ emission and limited sky coverage of other optical emission lines at very high sensitivity and velocity resolution, but poor spatial resolution \citep[a 1\degr\ beam;][]{haffner03,whampy}. X-ray coverage, tracing the hot ionized gas, was mapped earlier with ROSAT \citep{Motch1991} and more recently expanded with eRosita \citep{Predehl2021}. 

In the Magellanic Clouds, the Spizter and Herschel Space Telescopes surveyed the LMC and SMC \citep{SAGELMC,SAGESMC,HERITAGE} in the IR, while \citet{MAGMA} and \citet{MAGMA-SMC} provide CO maps at 45\arcsec\ beam size. In HI, \citet{Kim2003} provide combined ATCA and Parks data in the LMC at 1\arcmin\ resolution, while \citet{Stanimirovic1999} provides data on the SMC with 1.5\arcmin\ resolution. The SMC has also been observed by \citet{McClure2018} with ASKAP on 30\arcsec\ spatial scales. In the optical, \Halpha\ maps from WHAM provide a large scale picture of the low density warm ionized gas in the LMC \citep{Smart2023}, SMC \citep{Smart2019}, and Magellanic Bridge \citep{Barger2013}. Higher resolution narrow-band imaging has also been undertaken for a limited set of strong emission lines \citep{MCELS}. Finally, with regards to low-metallicity massive stars, the recently-completed {\it Hubble Space Telescope} Director's Discretionary program, Ultraviolet Legacy Library of Young Stars as Essential Standards (ULLYSES, \citealt{Roman-Duval2020}),
 and optical follow-up, XShooting-ULLYSES (XShootU, \citealt{Vink2023}), are providing UV to NIR spectra, stellar and stellar-wind parameters, and ionizing fluxes, which are crucial for some comprehensive LVM studies. These two highly-complementary spectroscopic surveys targeted $\sim250$ stars in the LMC and SMC (mostly) and a couple of stars in NGC 3109 and Sextans A. In the LMC and SMC, the surveys uniformly sample the fundamental astrophysical parameter space of high-mass stars while in the other two galaxies they provide important clues about stars at even lower (one tenth solar) metallicity.

The availability of these diverse datasets, as well as many others not listed here, make a wide-area optical spectroscopic survey with LVM's scope particularly timely.

\subsection{Survey Definition}
We now define the survey to be conducted within the LVM project that drives the design of the hardware and survey infrastructure. Fig.~\ref{fig:overview} shows the planned footprint of the survey and visualizes the spatial resolution achieved in the Milky Way and in the Magellanic System. An overview of the LVM targets is given in Table~\ref{tab:targets}. LVM cannot avoid the trade-off between sky coverage and spatial resolution. 
A reasonable balance is reached by aiming for resolving the internal structure of individual star-forming \HII\ regions, separating the gas affected by individual sources of ionization (massive stars) in the Milky Way ($<$1\,pc scales), while reaching sufficient resolution in the Magellanic Clouds to study gradients in physical conditions across \HII\ regions, and to resolve the largest young stellar clusters (10\,pc scales).

{\bf Milky Way Survey:} covers the bulk of the MW disk observable from the southern hemisphere, including the vast majority of known optically selected \Hii\ regions and their interface with the diffuse ionized ISM. Our target line depth will allow us detect strong emission lines to map the metallicity and ionization structure of the ISM over the full surveyed area, characterize the feedback-induced kinematics of gas in and around \Hii\ regions, and study electron temperature fluctuations (via auroral lines) and their impact on chemical abundance measurements for a significant fraction of optically visible \Hii\ regions in the sky. Sub-parsec resolution while covering kpc to global galactic scales allows for feedback effects to be traced starting from the individual sources through ionization fronts and shocks to global effects such as gas flows and galactic fountains.

{\bf Magellanic System Survey:} covers the nearby metal-poor SMC and LMC to a line sensitivity limit allowing for the detection of strong emission lines over the full area and auroral lines in the majority of \HII\ regions, sampling a lower metal abundance regime than is available in the MW while still resolving the inner structures of \HII\ regions. At the same time, the measurement of the stellar continuum will enable an unprecedented picture of metal-poor stellar populations. Assuming past average weather conditions, we can reach 100\% of D$_{25}$ in the SMC and LMC within 4 years. At 10\,pc resolution, we will distinguish shocked vs.\ photoionized emission, resolve large scale ionization gradients in star-forming regions, and fully sample all sites of star formation and large scale diffuse gas globally.

{\bf Nearby Galaxies Survey:} covers nearby ($<20$\,Mpc) galaxies when other targets are not observable on a best-effort basis. This sample of galaxies provides data on many galaxies with large angular extent in the nearby universe that cannot be covered with existing IFUs. It provides a dataset bridging the LVM and more traditional extragalactic IFU surveys. Some of these galaxies have metallicities below that of the SMC. They are not part of the core survey and hence do not drive requirements.
\section{Science Goals and Requirements}\label{sec:requirements}

In this section, we discuss the LVM survey science goals and how these translate into the science requirements that drive the design of the LVM-I and the experimental design of the survey. This list is by no means a comprehensive set of science cases that can be pursued using the LVM data. They are chosen to bracket observables likely to be key for a wide range of science projects yet simple enough to simulate and analyze to derive requirements on survey and instrument. In each subsection, we briefly discuss the scientific goals the LVM aims to tackle, and the characteristics that the LVM data must have in order to perform the measurements required.

\subsection{Quantifying the injection of feedback into the ISM around individual massive stars in the MW and young clusters in the Magellanic clouds.}

The LVM will map the physical conditions (kinematics, temperature, density, and ionization state) of ionized gas in the vicinity of young, high mass stars and young clusters. Quality stellar spectroscopy is now widely available, providing quantitative estimates of hot star luminosities, effective temperatures, and metallicities, and hence measures of ionizing spectra and wind properties \citep[see, e.g.,][]{Zari2021}. The combination of gas and stellar data will allow us to study quantitatively how energy and momentum is deposited into the surrounding ISM, driving turbulence, bulk motions of gas, and heating the gas via photo-ionization and shocks. Even in cases when individual star spectroscopy is not possible in the Magellanic Clouds, young cluster properties can be derived from Color Magnitude Diagrams \citep{bitsakis17}, adding an outside, nearly face-on perspective of feedback in a lower metallicity, HI-dominated environment.

\subsubsection{Maps of the line-of-sight velocity of ionized gas}\label{sec:losvd}

Small scale ($<$0.1\,pc) bulk gas motions in MW \HII\ regions are observed to be of the order of $\sim 5$\,km\,s$^{-1}$ \citep{lagrois09}. On somewhat larger scales ($>$1\,pc), bulk flows of $\sim10-50$\,km\,s$^{-1}$ are observed in the MW \citep[e.g.,][]{hausen02, russeil16} and in the LMC/SMC \citep[e.g.,][]{chu94}. 

To observe these motions we require spectra of sufficient spectral resolution and S/N to ensure a $3\,\sigma$ measurement of $v_{\rm gas}=5$\,km\,s$^{-1}$. The maps should sample \HII\ regions below $\sim 1$\,pc scales to separate the areas of influence of individual ionizing stars in young massive clusters and OB associations \citep{portegies10} in the MW, and at $\sim 10$\,pc scales to separate the areas of influence of individual young ionizing clusters in the MCs \citep{Glatt10}. Additionally, a large dynamic range in \HII\ region properties (e.g., luminosity, age, metallicity, morphology, environment) is desired to study how the feedback process depends on local conditions.

The DESI spectrograph design adopted by LVM delivers an approximately Gaussian line-spread-function (LSF) with $\sigma_{\rm ins}\simeq30$\,km\,s$^{-1}$ ($R\simeq$4000 at 6563\,\AA). Simulations of the centroiding error for single emission lines at this resolution show that an emission line flux S/N of 30 (20) is necessary to reach a $3\sigma$ precision of 5\,km\,s$^{-1}$ (10\,km\,s$^{-1}$). Typically, information from multiple lines will be available, but different ions might trace gas in structures with different kinematics (e.g., shock and ionization fronts, or the surfaces of dense gas clumps), so we take the conservative approach of assuming the fiducial $v_{\rm gas}$ measurement is done solely from \Halpha. 

In their fainter, outer parts where the transition to the WIM occurs, MW and LMC/SMC \HII\ regions have typical \Halpha\ surface brightnesses of $\sim6\times10^{-17}$\,erg\,s$^{-1}$\,cm$^{-2}$\,arcsec$^{-2}$ \citep{Haffner2009,Krishnarao2017}. Reaching a line flux S/N=30 (20) in these regions therefore requires reaching a $5\sigma$ line depth of $1\times10^{-17}$ ($1.5\times10^{-17}$) \,erg\,s$^{-1}$\,cm$^{-2}$\,arcsec$^{-2}$.
Spatially resolving $<1$\,pc sized structures out to a distance of 3\,kpc \citep[which encompassses more than 2/3 of known optically selected \HII\ regions,][]{blitz82} in the MW, and 10\,pc at the distance of the LMC requires an instrument with $<40$\arcsec\ spatial resolution. 

The 3\,kpc distance limit matches the distance distribution of young and massive stellar targets in the MWM program \citep{Zari2021}. Out to this distance we can assume a typical foreground dust extinction of $A_{\rm V}<2.2$\,mag, given by the third quartile of integrated extinction towards typical regions of interest using the MW 3D dust maps of \citet{green15} (see also Section~\ref{sec:mwsel} below).

\subsubsection{Maps of emission line velocity dispersion}

The electron temperature in \HII\ regions in the solar neighborhood is $\sim8,000$\,K \citep{quireza06}, implying a thermal line width floor in Hydrogen for individual velocity components of $12$\,km\,s$^{-1}$. In the LMC and SMC, temperatures are higher, $\sim10,000$\,K and $\sim12,000$, respectively \citep{tsamis03}, implying a $15$\,km\,s$^{-1}$ floor. The observed velocity profiles will have contributions from multiple velocity components along the line of sight and from turbulent broadening. To characterize turbulent motions, bulk gas motions, and separate velocity components to the best extent possible at the 30\,km\,s$^{-1}$ instrumental resolution of the LVM, we require the ability to obtain a $3\sigma$ measurement of the thermal line width floor.

Simulations of the line width error achievable for single emission lines with the LVM spectrograph \citep[see also][for state-of-the-art LSF modeling]{law21}
show that a S/N$=40$ (30) is necessary to reach a $3\sigma$ precision of $12$\,km\,s$^{-1}$ ($15$\,km\,s$^{-1}$), assuming perfect knowledge of the line-spread-function (LSF). Introducing a 5\% error on the knowledge of the LSF degrades the precision to $18$\,km\,s$^{-1}$ ($20$\,km\,s$^{-1}$). Using the same \Halpha\ surface brightness limit assumed in Section~\ref{sec:losvd}, we require reaching a $5\sigma$ line depth of $\simeq8\times10^{-18}$\,erg\,s$^{-1}$\,cm$^{-2}$\,arcsec$^{-2}$ in the MW, and $1\times10^{-17}$\,erg\,s$^{-1}$\,cm$^{-2}$\,arcsec$^{-2}$ in the MCs, with spatial resolution requirements being the same.

\subsubsection{Maps of strong emission line ionization parameter and shock diagnostics}\label{sec:mwsel}

Strong emission line (SEL) ratios can be used to measure the ionization state of the gas, and also provide evidence for the presence of shocks.  For example, [OIII]$\lambda\lambda$4959,5007/[OII]$\lambda\lambda$3726,3729 and [SIII]$\lambda\lambda$9069,9532/[SII] $\lambda\lambda$6717,6731 are good diagnostics of ionization parameter \citep{Kewley2019}. Similarly, different forms of ionization can be probed by using the classical emission line ration diagnostics, [OIII]/Hb versus [NII]/Ha \citep{Baldwin1981} and [OIII]/Hb versus [SII]/Ha \citep{veilleux1987}. We require a detection of all these emission lines across the regions of interest.

Adopting a typical dust extinction of $A_V=3.4$ mag (2.2\,mag of foreground extinction as discussed in Section \ref{sec:losvd} plus 1.2\,mag of intrinsic extinction, 
\citealp{Cox1987}), and a fiducial reddened low excitation CLOUDY photo-ionization model with solar metallicity (typical of the outskirts of MW \HII\ regions), the lowest SEL ratio relative to \Halpha\ is expected to be \oii$\lambda\lambda$3726,3729/\Halpha\ $\simeq 0.1$.
This, combined with the lower DESI spectrograph system throughput in the blue, implies that the strongest requirement for SEL detection in the MW is set by the ability to detect the \oii\ doublet.  

Using the same fiducial \Halpha\ surface brightness at the outskirts of MW \HII\ regions as in Section \ref{sec:losvd},
an observed \oii/\Halpha\ ratio of 0.1, and the DESI spectrographs system throughput curve, we calculate that a $3\sigma$ detection of \oii\ requires reaching a $5\sigma$ flux limit of $\sim8\times10^{-18}$\,erg\,s$^{-1}$\,cm$^{-2}$\,arcsec$^{-2}$ in \Halpha. 

For a typical level of dust extinction towards LMC/SMC \HII\ regions of $A_V=1.0$ mag \citep[$<0.3$\,mag of foreground extinction plus 0.7\,mag of intrinsic extinction,][]{Caplan1986,Caplan1996,misselt99}, the lowest line ratios relative to \Halpha\ predicted by a fiducial reddened, low excitation, 40\% solar metallicity CLOUDY photo-ionization model are $\simeq0.05$ for \nii$\lambda$6583/\Halpha, and \sii$\lambda$6717/\Halpha. Due to the lower metallicity and dust extinction of the gas with respect to the MW, for the LMC and SMC the strongest requirement for SEL detection is therefore set by the ability to detect \nii\ and \sii. 
An analogous calculation to the one above implies a required $5\sigma$ surface brightness depth of $\sim5\times10^{-18}$\,erg\,s$^{-1}$\,cm$^{-2}$\,arcsec$^{-2}$ in \Halpha\ in order to detect all SELs of interest in the MCs.

Characterizing the physical state (e.g., density, velocity, magnetic field strength) of shocked gas, and mapping the ionization state of the gas across the whole nebula requires comparing the observed line ratios to the predictions of photo-ionization models. These models suffer from systematic uncertainties at the 10-20\% level in the predicted line fluxes \citep[e.g.,][]{Ji2020,Ballhausen2023}. We therefore require the relative flux calibration errors across the full LVM wavelength range to be smaller than 10\% so that calibration errors are not the dominant source of error.

\subsection{Directly observe the origin of the abundance discrepancy problem and the biases affecting SEL abundance diagnostics.}
\label{sec:adf}

The LVM will produce the ideal dataset to solve the ``abundance discrepancy problem'' currently plaguing the derivation of nebular chemical abundances using different methods \citep{Garcia-Rojas2007, kewley08, blanc15, blanc19, MendesDelgado2023a}. It is well established that nebular abundances derived using the direct $T_e$ method, SEL methods calibrated against theoretical photo-ionization models, and methods based on the direct measurement of metal recombination lines (RL), show systematic discrepancies at the 0.2-0.3\,dex level and sometimes larger. The inability to measure gas-phase chemical abundances to better than a factor of two is a severe limiting factor in our understanding of the chemical evolution of galaxies and the Universe \cite[e.g.][]{lu15}. Several origins have been proposed for these discrepancies, including temperature and density inhomogeneities in ionized nebulae \citep{Peimbert1967,MendesDelgado2023a,MendesDelgado2023b}, non-Maxwellian electron temperature distributions \citep{Nicholls2012}, local metallicity variations \citep{Torres1990,Liu2000}, and assumptions about the ionization structure and geometry of ionized nebulae. For bright \HII\ regions in the MW, the LVM will map both strong nebular metal lines and temperature-sensitive auroral lines, producing spatially resolved maps of chemical abundances using different methods, while simultaneously mapping the geometry and ionization state of the gas. For the same regions, the MWM will produce photospheric stellar abundances of the same elements \citep[O, N, S, Ar, see, e.g.,][]{izotov06} for several of co-spatial young massive stars, which provides a unique reference to evaluate and solve the systematics affecting different abundance diagnostics.

\subsubsection{Maps of abundance and electron density diagnostic SEL ratios across nearby bright MW \HII\ regions}\label{sec:sel-abundances}

The gas-phase metallicity (or oxygen abundance) can be estimated using the indirect method based on the strong emission line ratios in cases where the intrinsically faint auroral lines can not be detected with enough S/N needed to use the more robust direct T$_e$ method (see Section \ref{sec:auroral}). The indirect methods are empirical or theoretical model-based relation between the gas-phase metallicities and the strong line ratios such as R23 \citep[=\oii$\lambda\lambda$ 3726, 3729 + \oiii$\lambda\lambda$4959, 5007/H$\beta$;][]{Pagel1979}, N2 \citep[=\nii$\lambda$6583/H$\alpha$;][]{Denicolo2002}, O3N2  \citep[= \oiii$\lambda$ 5007/H$\beta$/\nii$\lambda$6583/H$\alpha$;][]{Pettini2004}. We refer readers to \citet{Maiolino2019} for a detailed review of the indirect methods. The electron density of an \HII\ region can be estimated from the density-sensitive emission line flux ratios such as \oii$\lambda\lambda$3726,3729, H$\beta$ and \sii$\lambda\lambda$6717,6731. \citep{osterbrock89}.

Hence, to estimate gas-phase metallicities and electron densities, we must obtain extinction corrected (via Balmer/Paschen decrement) maps of the following transitions: \oii$\lambda\lambda$3726,3729, H$\beta$, \oiii$\lambda\lambda$4959,5007, \Halpha\, \nii$\lambda\lambda$6548,6583, \sii$\lambda\lambda$6717,6731, [ArIII]$\lambda\lambda$7135,7751, and \siii$\lambda\lambda$9069,9532, across a sample of bright nearby \HII\ regions in the MW. We require a detection of all these emission lines across the regions of interest. The emission line maps should spatially resolve \HII\ regions below $\sim 1$\,pc scales to separate the areas of influence of individual ionizing sources \citep{portegies10}. 
In addition, the LVM spectral resolution should be sufficient to resolve emission line doublets, including the 3\,\AA\ separation of the \oii\ doublet which is a good electron density diagnostic. Spatial resolution requirements are the same as in Section~\ref{sec:losvd}.

Assuming the same total dust extinction as in Section~\ref{sec:mwsel} ($A_V=3.4$\,mag), and a fiducial solar metallicity intermediate ionization CLOUDY photo-ionization model, the lowest line ratio relative to \Halpha\ is \oii$\lambda$3727/\Halpha=0.05. Assuming an \Halpha\ surface brightness of interest of $\sim2\times10^{-15}$\,erg\,s$^{-1}$\,cm$^{-2}$\,arcsec$^{-2}$, which is representative of bright optically selected \HII\ regions, an observed \oii/\Halpha\ ratio of 0.05, and the DESI spectrograph system throughput curve, a $3\sigma$ detection of \oii\ requires reaching a $5\sigma$ \Halpha\ flux limit of $\sim4\times10^{-17}$\,erg\,s$^{-1}$\,cm$^{-2}$\,arcsec$^{-2}$. 

Typical systematic uncertainties in the emissivities of these transitions, associated with systematic uncertainties in the determination of the gas physical conditions and by the accuracy of available atomic data are at the 5-10\% level \citep{Ballhausen2023}. We therefore require the relative flux calibration errors to be smaller than 5\% from 3700\,\AA\ to 9600\,\AA\ so calibration uncertainty is not the dominant source of error in abundance determinations.

\subsubsection{Maps of temperature sensitive auroral lines of different ions across bright optically detected MW \HII\ regions} 
\label{sec:auroral}

Auroral lines in combination with strong nebular lines provide a direct measure of the electron temperature, and therefore of the collisionally excited line emissivities. Observing these transitions is therefore essential in order to measure ionic and elemental abundances using collisionally excited lines. Because of the high metallicity and associated low electron temperatures in MW \HII\ regions, auroral lines have relatively low emissivities. Their intrinsic faintness combined with the high levels of dust extinction expected in the MW plane makes auroral line detection challenging. 

The LVM baseline is optimized to produce maps of the auroral lines \siii$\lambda$6312 and \oii$\lambda$7320 for nearby ($<3$\,kpc) low-extinction ($A_V<2.0$) \HII\ regions. These transitions trace the high and low ionization zones of the nebulae, and their redder wavelengths make them less subject to dust extinction than other commonly used lines like \oiii$\lambda$4363 and \nii$\lambda$5755. The latter transitions will be detectable in very bright and/or low-extinction regions, but the survey is not designed to ensure their detection in all areas of interest.

Using the same reference photo-ionization model (of solar metallicity)  as in the previous section, but reddened by $A_V=2.0$\,mag of extinction only, we expect to observe line ratios with respect to \Halpha\ of \siii$\lambda$6312/\Halpha=0.005 and \oii$\lambda$7320/\Halpha=0.003 (for reference a ten times lower line ratio of 0.0004 is expected for \oiii$\lambda$4363, mostly due to extinction). As before, for bright \HII\ regions we consider an \Halpha\ surface brightness of $\sim2\times10^{-15}$\,erg\,s$^{-1}$\,cm$^{-2}$\,arcsec$^{-2}$. Considering the DESI spectrograph throughput curve we calculate that a $3\sigma$ detection of \oii$\lambda$7320 requires us to reach a $5\sigma$ flux limit of $\sim1\times10^{-17}$\,erg\,s$^{-1}$\,cm$^{-2}$\,arcsec$^{-2}$ in \Halpha. To ensure calibration errors on the \oii$\lambda$7320/\oii$\lambda$3727 ratio are sub-dominant compared to systematic errors in abundance determinations, we require $<5\%$ relative flux calibration errors from 3727\,\AA\ to 7320\,\AA.

\subsection{Characterize the dispersal and large-scale distribution of metals in the Magellanic Clouds}

The LVM will provide measurements of \HII\ region chemical abundances across our targets. While the level of spatial detail will be 10 to 100 times lower than in the MW, the SMC and LMC provide a large dynamic range in metallicity when combined with the MW. Also, the external view of these targets allows us to map the chemical composition of ionized nebulae across their disks, without the challenges that dust extinction and distance determinations impose on MW studies. To do this we need to measure SEL abundance diagnostics across these targets, and auroral lines where possible. Also, to trace the dispersal of chemical elements into the ISM we push the survey depth to map these emission lines out to the transition into the WIM, beyond the boundary of the discrete ionized nebulae.

\subsubsection{Maps of SEL abundance diagnostics across the LMC and SMC}

We need to measure the same transitions as in Section~\ref{sec:sel-abundances}. Due to the lower metallicity and higher excitation of the gas with respect to the MW \citep{russell92, pellegrini12}, for the LMC and SMC the strongest requirement for SEL detection is set by the ability to detect \nii$\lambda$6583 and \sii$\lambda$6717, for which we expect line ratios of $\sim0.05$ relative to \Halpha\ (see Section~\ref{sec:mwsel}). Adopting a typical \Halpha\ surface brightness at the fainter, outer parts of \HII\ regions where the transition to the WIM starts of $\sim6\times10^{-17}$\,erg\,s$^{-1}$\,cm$^{-2}$\,arcsec$^{-2}$, and the DESI spectrograph system throughput curve implies that a $3\sigma$ detection of these lines requires reaching an \Halpha\ $5\sigma$ flux limit of $\sim5\times10^{-18}$\,erg\,s$^{-1}$\,cm$^{-2}$\,arcsec$^{-2}$.

\subsubsection{Maps of auroral lines across the LMC and SMC}

As discussed in Section \ref{sec:auroral}, auroral lines permit a direct measure of the electron temperature, and from it, of the ionic and elemental chemical abundance. Observing these transitions in the LMC and SMC will help us characterize the chemical abundance of different elements across these galaxies, controlling for the systematics of SEL diagnostics and their possible dependence on environment. It will also help us understand how and why the abundance discrepancy problem depends on metallicity (see Section \ref{sec:adf}).

Because of the low metallicity and higher electron temperatures in the Clouds, auroral lines have relatively high emissivities compared to the MW. For an average $A_V=1.0$\,mag level of dust extinction towards LMC/SMC HII regions (Section \ref{sec:mwsel}), and the reference reddened low metallicity (40\% solar) CLOUDY photo-ionization model assumed in Section \ref{sec:mwsel}, we predict auroral line ratios relative to \Halpha\ of 0.06, 0.08, and 0.05 for the \oiii$\lambda$4363, \siii$\lambda$6312 and \oii$\lambda$7320 transitions, respectively. We do not aim to detect auroral lines all the way out to the WIM transition. The lowest \Halpha\ surface brightness of interest for discrete \HII\ regions in the LMC/SMC is $\sim2\times10^{-16}$\,erg\,s$^{-1}$\,cm$^{-2}$\,arcsec$^{-2}$. For a line ratio of 0.05, a $3\sigma$ detection of \oii$\lambda$7320 requires reaching $5\sigma$ flux limit of $\sim2\times10^{-18}$\,erg\,s$^{-1}$\,cm$^{-2}$\,arcsec$^{-2}$ at the wavelength of \Halpha. Given the above line ratios and the shape of the DESI spectrograph throughput curve, a detection of \oii$\lambda\lambda$7320,7330 ensures a detection of \siii$\lambda$6312. \oiii$\lambda$4363 will be detected wherever the brightness allows.

\subsection{Observe and quantify the heating mechanisms in the DIG and the escape of UV radiation}

Existing wide field maps of the MW ionized ISM, like the Wisconsin \Halpha\ Mapper (WHAM) \citep{haffner03}, or the combined MW \Halpha\ map presented in \citet{fink03}, show that faint MW \Halpha\ emission from the DIG is present everywhere on the celestial sphere, and that it contains significant substructure.  WHAM has a beam of $\sim 1^\circ$ and the \citet{fink03} map, which combines WHAM data with Virginia Tech Spectral line Survey \citep[VTSS,][]{dennison98}) data in the north, and the Southern \Halpha\ Sky Survey Atlas \citep[SHASSA,][]{gaustad01} in the south, resolves between 6' and $1^\circ$. 

LVM is designed to map large portions of the DIG in the MW for bright emission lines in the full optical range (from \oii$\lambda$3727\AA\ to \siii$\lambda$9532\AA) at $<1$ arcminute scales. Aided by spatial co-adding, fainter MW DIG emission lines like HeI$\lambda$5876 and \siii$\lambda\lambda$9069,9531 can be detected on larger spatial scales ($\sim$10-100'). These fainter lines allow better determinations of temperature and constrain the amount of leaky photons from \HII\ regions needed to photoionize the DIG \citep[e.g.,][]{Haffner2009,hoopes03}.
At the outer-edge of star forming regions, the nebular lines from low ionization species reach a surface brightness of the order of the \Halpha\ intensity. These are enhanced over canonical values of integrated nebulae because the volume of gas where low ionization emission is emitted in optically thick \HII\ regions is significantly smaller than that of the fully ionized gas, making it easier to detect. This transition is characterized by large changes in both the absolute surface brightness and also the relative intensities of emission lines.
 
Studies of the DIG in the MW and other galaxies \citep[e.g.,][]{zhang17,jones17} revealed a possible need for a heating source other than photoionization from \HII\ regions to consistently explain the emission lines \citep[e.g.,][]{Haffner2009}.  The increase in \nii/\Halpha\, \sii/\Halpha\, and \oiii/H$\beta$ with decreasing \Halpha\ surface brightness, while \nii/\sii\ remains constant, is difficult to reproduce with models of photoionization from leaky \HII\ regions alone \citep{hoopes03,barnes14,belfiore22}, though 3-D photoionization \citep{Wood2005,Wood2013,Wood2019} or self-consistent hydro simulations with ray tracing \citep{Kado-Fong2020,Kim2023} show it is possible.
The spatial resolution and wealth of spectral information of the LVM data will allow us to correlate the observed diffuse structures and their emission line ratios with nearby \HII\ regions and other possible additional heating sources.

The distinction between DIG and dense \HII\ regions also delineates the sphere of influence of feedback from stars and massive clusters. The same line ratios that probe DIG ionization reveal leaky \HII\ regions. On scales of $\sim 10 - 50$\,pc, relatively dense \HII\ regions surrounding massive compact clusters reprocess much of the emitted ionizing stellar radiation, as evidenced by their large contribution to the integrated \Halpha\ luminosity of typical galaxies. These dense nebulae absorb stellar photons and prevent them from heating the large scale diffuse ISM in galaxies. Theoretical studies suggest the barrier is porous \citep{howard2017}  which can explain observations indicating high leakage \citep[e.g.,][]{Haffner2009, seon2009} which vary with environment or with cluster luminosity \citep[e.g.,][]{beckman2000}. LVM will spatially resolve the transition from the interiors of star-forming regions into their surrounding ISM. This will provide two critical diagnostics of the escape fraction of UV continuum: the extinction-corrected \Halpha\ luminosity and an ionization parameter map \citep[][]{pellegrini12}. These two constraints allow for the determination of the covering fraction of escaping radiation.

\subsubsection{Maps of SELs in High Galactic Latitude Fields}

The LVM data must trace strong emission lines used to characterize the DIG with sufficient S/N to compare to photo-ionization models. The survey must target areas of diffuse emission from warm ionized gas, which can have scale heights of up to 1\,kpc in the MW, and a surface brightness at these scale-heights of $\sim 10^{-18}$\,erg\,s$^{-1}$\,cm$^{-2}$\,arcsec$^{-2}$ \citep{Krishnarao2017}. We need to observe regions where the DIG is not heavily contaminated by other nebular emission sources, which requires focusing on regions above the plane of the MW disk ($|b|\gtrsim 5-30^\circ$). 

For the expected line ratios relative to \Halpha\ in the DIG of 0.3-0.5 for \sii$\lambda$6717 and \nii$\lambda$6584, the LVM spectra should reach a S/N=10 in \Halpha\ in order to have significant (3-5 $\sigma$) detections of strong low ionization nebular lines. Resolving the spatial structure of the MW DIG to a factor of several better in spatial resolution than current studies \citep[e.g., WHAM][, with $\sim1^\circ$ resolution]{haffner03} requires reaching this S/N on scales of $\sim 10'$. For the fiducial 37\arcsec\ spaxel size of the LVM system, a beam size of 9\arcmin\ is reached after binning $15\times15$ spaxels. This translates into a requirement to reach a $5\,\sigma$ limiting depth per spaxel of $7\times10^{-18}$\,erg\,s$^{-1}$\,cm$^{-1}$\,arcsec$^{-2}$ at \Halpha.

\subsubsection{Extinction-Corrected \Halpha\ Luminosities of MW and LMC/SMC \HII\ Regions}

Ionizing photon escape fractions are derived by comparing the predicted Lyman continuum UV luminosity of the ionizing massive stars or young clusters, to the total hydrogen recombination rate derived from the observed \HII\ region \Halpha\ luminosity after correcting it for dust extinction (typically using the Balmer decrement method). The typical systematic uncertainty in the ionizing luminosity of different stellar atmospheres is of the order of 50\% for state-of-the-art stellar atmospheres and stellar population models, depending on the spectral type and assumptions regarding the role of rotation and binarity in high mass stellar evolution. We therefore require an uncertainty on the de-reddened \Halpha\ luminosity of 20\% (S/N=5) to keep the LVM data from being the dominant source of uncertainty in these studies. 

For an average expected level of dust extinction towards optically detected $D<3$\,kpc MW \HII\ regions ($A_{\rm V}=3.4$ mag, see Section \ref{sec:mwsel}), we employ reddening simulations to estimate the line flux depth needed to recover a final S/N of 5 in the Balmer decrement extinction-corrected \Halpha\ luminosity, under the conservative assumption that only the \Halpha/H$\beta$ ratios is used to calculate the dust extinction. We assume an intrinsic ratio \Halpha/H$\beta$=2.86 \citep{osterbrock89} and a MW ($R_V=3.1$) extinction law. Using the \citet{fink03} map of MW \Halpha\ emission we estimate a surface brightness of interest limit of $1\times10^{-16}$\,erg\,s$^{-1}$\,cm$^{-2}$\,arcsec$^{-2}$, which encompasses most discrete \HII\ regions seen in the map and recovers $>90\%$ of the total \Halpha\ flux of the Galaxy. Since the total luminosity of HII regions is computed by integrating over large areas on the sky, the dominant source of error comes from the dust corrections. At this observed \Halpha\ surface brightness, the assumed level of extinction implies an observed \Halpha/H$\beta$ ratio of 9.5, and a S/N=11 requirement on the observed H$\beta$ flux to keep extinction correction errors below the 20\% level. Assuming a modest integration area of $6\times6$ spaxels ($\sim$4') for barely resolved HII regions, this requirement is met by reaching a $5\,\sigma$ limiting depth per spaxel of $3\times10^{-17}$\,erg\,s$^{-1}$\,cm$^{-2}$\,arcsec$^{-2}$ at \Halpha. 

Relative calibration errors $<5\%$  from 6563\,\AA\ to 4861\,\AA\ are necessary to ensure $<10$\% bias on the dereddened \Halpha\ flux due to bias in the Balmer decrement. This is adequate given our S/N=5 goal in dereddened \Halpha\ flux. Errors below $<10\%$ in absolute flux are needed to ensure uncertainties of the total ionizing flux necessary to account for the energy budget ionizing the DIG are sub-dominant to uncertainties in FUV output of massive stars.

The study of the escape of ionizing radiation from dense \HII\ regions in the LMC and SMC offers an external view. While at lower resolution than the MW, the reduced attenuation by dust, as well as an increase in the mean free path of radiation due to lower metal and dust abundances, allow for more accurate studies of the escape fraction. LVM will make it possible to compare the \Halpha\ fluxes of the \HII\ regions with that of the DIG and  the expected Lyman continuum luminosity of their ionizing stellar populations, in order to estimate the escape fraction and the mean free path of Lyman continuum photons. Further, the addition of SELs will allow for the spectral shape of the ionizing radiation field to be compared with that of nearby clusters. Finally, an accurate measurement of the total integrated emission, from both \HII\ regions and DIG, can be compared with the escape luminosities derived from individual sources, where the difference is the galactic escape fraction. Current estimates place the SMC galactic escape fraction of ionizing radiation at 4-9\%, and 10-20\% for the LMC. \citet{Barger2017} conducted a study of escaping ionizing radiation from the Magellanic Clouds and Milky Way using diffuse \Halpha\ maps from WHAM, which have informed many models of the large scale radiation field environment of the Milky Way and Magellanic Clouds \citep{Bland-Hawthorn2019,Antwi-Danso2020}.

For the typical expected level of dust extinction of LMC/SMC \HII\ regions of $A_{\rm V}=1.0$ mag, reddening simulations indicate that a S/N=12 is required in the H$\beta$ flux to ensure a S/N=5 measurement of the Balmer decrement extinction corrected \Halpha\ luminosity of HII regions. Using the MCELS \citep{MCELS} narrow-band \Halpha\ maps, we conclude that a surface brightness of $2\times10^{-16}$\,erg\,s$^{-1}$\,cm$^{-2}$\,arcsec$^{-2}$ encompasses most discrete \HII\ regions. We therefore require reaching a $5\sigma$ limiting depth of $2\times10^{-17}$\,erg\,s$^{-1}$\,cm$^{-2}$\,arcsec$^{-2}$.  This is a per spaxel depth and assuming no spatial binning given the lower resolution of the LVM on the Magellanic Clouds. With this limit we recover $>90\%$ of the total \Halpha\ flux in the MCs. 
	
\subsection{Stress-test and Calibration of Stellar Population Synthesis (SPS) models for young stellar clusters and field stars}

The LVM will be particularly powerful at obtaining integrated spectroscopy of young stellar clusters in the Magellanic Clouds, and of MW globular clusters. The external view of the former, and compactness of the later minimizes the effects of foreground and background contamination from field stellar populations. Of particular interest are clusters with available color-magnitude diagrams (CMDs) from e.g. the Hubble Space Telescope \citep{sarajedini+07,milone+23}, the VISCACHA survey \citep{maia+19,dias+20} and others. Many degeneracies associated with the stellar population modeling of mixed stellar populations are broken when analyzing individual stellar clusters, providing further opportunities for refining and calibration stellar population synthesis models (e.g.\ \citealp{schiavon+04,asad+20,asad+22}). Furthermore, the presence of massive stars in young clusters allows for the opportunity to constrain models of high mass stellar evolution and atmospheres by demanding simultaneous and self-consistent modeling of both the integrated spectra and the CMDs of these systems (e.g.\ \citealp{leitherer05,coelho+20}).

\subsubsection{Integrated Continuum Spectra of Young Clusters in the LMC, SMC}

In order to reach formal uncertainties of $\sim0.1$\,dex in mean stellar age, metallicity, and abundances via absorption index fitting in the LVM spectra we need to reach a continuum $V$-band S/N$\simeq 10$ per spaxel for regions with recent star formation \citep{kauffmann03} and intermediate-age star clusters, e.g.\ \citet{santos+04,piatti+05,dias+10}. These are conservative estimates as newly developed full spectral fitting techniques can lower these requirements significantly \citep[e.g.,][]{choi14}. We require the data cubes to have a V-band S/N$=12$ per \AA\, per spaxel. The data must separate individual clusters from the background stellar population on the targets. We must therefore resolve the $\sim 20$\,pc scales of individual young ionizing clusters. For a V-band surface brightness of $\simeq 22$\,AB\,mag\,arcsec$^{-2}$ \citep[typical of the outskirts of young clusters][]{carvalho08}, reaching a $V$-band S/N$\simeq 12$ per \AA\, per spaxel, requires reaching a $5\sigma$ $V$-band surface brightness limiting depth of $23$\,AB\,mag per \AA.

Eventually CMDs and integrated spectra should yield self-consistent results regarding the total stellar mass in a spatial region. Present day typical systematic uncertainty in stellar mass determination from spectral fitting is of 0.3\,dex or a factor of 2 \citep{kannappan07}. Therefore a 10\% calibration error in the spectra ensures this will be a sub-dominant source of error, even in face of improvements in the models which the LVM data might motivate. We also require Poisson limited sky subtraction over the range of 3700-9600\,\AA\ outside of skyline emission to ensure enough spectral information is available for stellar population fitting.

\subsection{Final Science Requirements}

Here we summarize the final science requirements defined as the most stringent requirements derived from the union of the key science cases described above. Table~\ref{tab:targets} provides an overview of the depth and resolution requirements by survey target, and Section~\ref{sec:lvmsurvey} describes the survey implementation in accordance with the requirements defined here.

{\bf Survey footprint:} The LVM survey footprint should encompass the bulk of all known $D<3$\,kpc, optically selected \HII\ regions in the MW visible from LCO. The footprint should also cover the full disks of the SMC and the LMC out to their optical radius. The surveyed areas should be based on a simple geometric selection function that allows straight-forward comparisons with theoretical models and numerical simulations. 

{\bf Emission line flux and continuum depth:} The survey spectra must reach an emission line $5\sigma$ depth of $6\times10^{-18}$\,erg\,s$^{-1}$\,cm$^{-2}$\,arcsec$^{-2}$ at 6563\,\AA\ in the Milky Way, and $2\times10^{-18}$\,erg\,s$^{-1}$\,cm$^{-2}$\,arcsec$^{-2}$ at 6563\,\AA\ in the Magellanic Clouds. Additionally, the SMC/LMC survey must reach a $5\sigma$ $V$-band continuum depth of 23.0\,AB\,mag\,arcsec$^{-2}$.

{\bf Spatial resolution: } The LVM data must have a spatial resolution better than $1$\,pc out to a distance of 3\,kpc in the MW (i.e. $<69''$), and sufficient to resolve typical \HII\ and young cluster scales of $<20$\,pc out to the $\sim63$\,kpc distance of the SMC (i.e. $<65''$). The LVM system optical design yields 37\,arcsec spaxels yielding a resolution of $\sim10$\,pc in LMC/SMC, and $<1$\,pc out to 6\,kpc within the MW.

{\bf Spectral resolution: } The system spectral resolution is largely set by the design of the adopted DESI spectrographs ($R \approx 4000$ at 6563\,\AA). This resolution is sufficient to satisfy all spectral resolution requirements for all the measurements described above.

{\bf Relative flux calibration: } We require a relative spectrophotometric calibration accuracy of better than 5\% over the wavelength range from (3700\,\AA-9600\,\AA) which covers all emission line and continuum features of interest. This allow us to satisfy all relative calibration requirements for all the measurements described above, including sub-dominant biases in Balmer decrement extinction corrections, electron temperature, and abundance determinations based on strong and auroral line ratios.

{\bf Absolute flux calibration: } We require an absolute spectrophotometric calibration accuracy of better than 5\%, which satisfies the requirements of all the science cases outlined above, and keeps calibration errors as a sub-dominant source of systematic error when comparing the LVM data to models of nebular photo-ionization and stellar population synthesis.

{\bf Sky Subtraction: } The goal for sky subtraction accuracy is Poisson limited performance everywhere. The requirement is to ensure systematic (i.e.\ non Poissonian) sky subtraction residuals that are small and random enough as to not significantly bias emission line measurements, and allow for S/N improvements when stacking spaxels. Based on the experience of the   SDSS-IV MaNGA survey we establish this as a requirement to achieve a ratio between sky subtraction residuals and expected Poisson residuals of $<1.3$ over the full wavelength range and $<1.1$ outside of bright sky emission lines \citep{2016AJ....152...83L}.

\section{The LVM Instrument (LVM-I)} \label{sec:instrument}

\begin{figure*}
\plotone{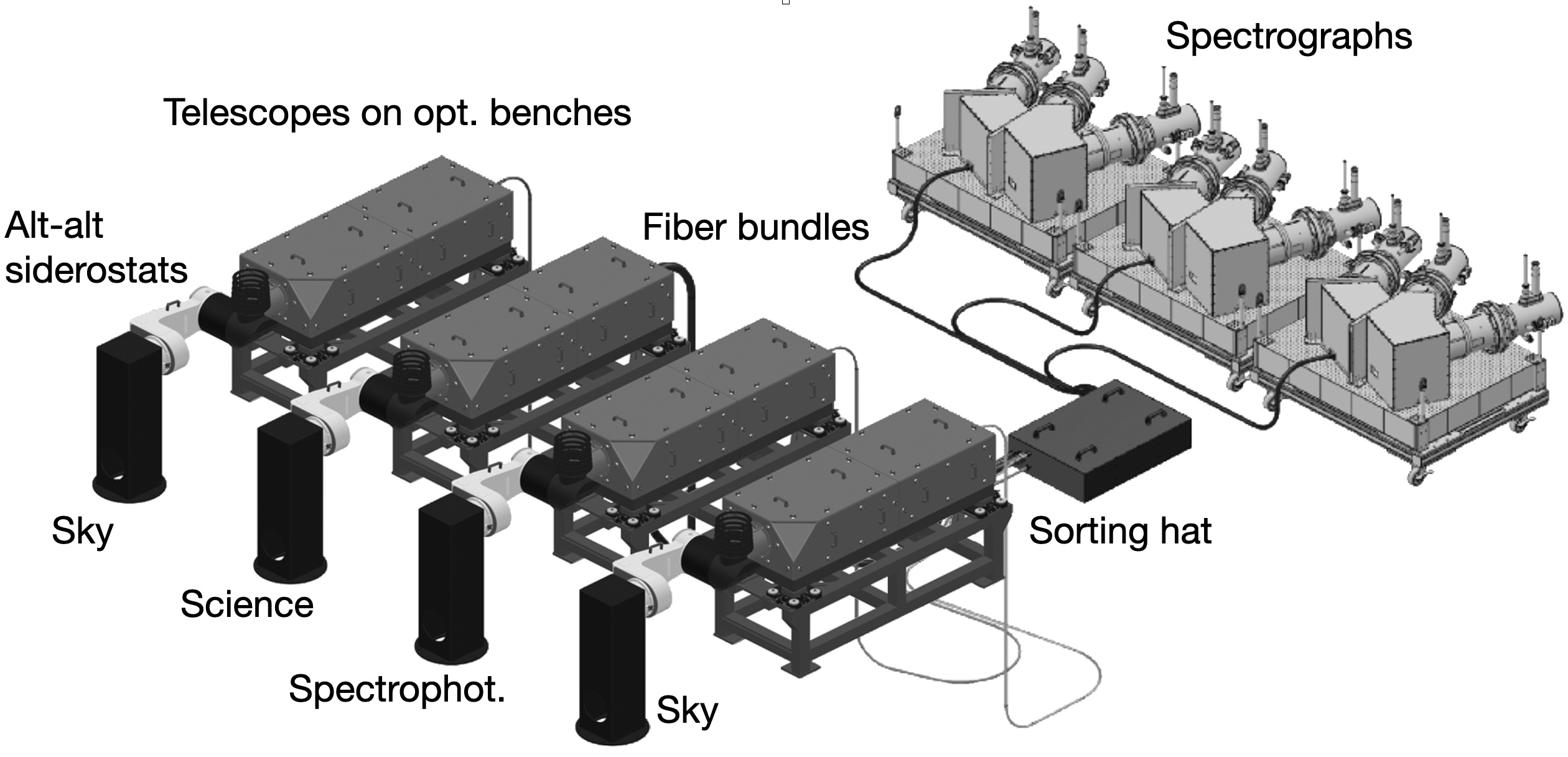}
\caption{The LVM-I instrument and its major subsystems.}\label{fig:system}
\end{figure*}

The LVM Instrument, or LVM-I, is a new observing facility designed specifically to provide the infrastructure needed to execute the LVM survey.
A full description of the LVM-I is given in \citet{konidaris24} and references therein \citep[see also][]{konidaris20}. The LVM-I is located at Carnegie's Las Campanas Obsevatory (LCO) in Chile. It is composed of four hardware subsystems: the telescopes \citep{Herbst20,Herbst22,Herbst2024}, the IFU and fiber system \citep{Feger20}, the spectrographs \citep{10.1117/12.2311996}, and the enclosure. Another major component is the instrument control and observing software described in detail in \citet{sanchez-gallego24}. Here, we will briefly describe the hardware systems necessary to understand LVM survey technicalities and the structure of the LVM data. An overview of the LVM-I labeling all major components is shown in Fig.~\ref{fig:overview}.

\subsection{Design Drivers and Constraints}

The LVM-I facility has a non-conventional, innovative design, that is driven by the requirements of the LVM program. In particular, the ultra-wide-field nature of the survey (thousands of square degrees) pushes for an order of magnitude increase in IFU field-of-view with respect to previous similar instruments. Mapping this large area in a reasonable amount of time requires moving from arc-minute sized IFUs (e.g.\ MUSE, MaNGA, PPAK, VIRUS-P) to fields of view spanning a significant fraction of a degree. Physical and budget limitations on the number and size of spectrographs and detectors that can be deployed, imposes a limit on the number of spaxels whose spectra can be imaged simultaneously. This limit in the number of spaxels pushes for large arcmin scale spaxels in order to reach the desired ultra-wide IFU FOV regime (instead of the arcsecond sized spaxels of the instruments mentioned above). Doing this, while at the same time reaching the ambitious physical spatial resolution scales required by the LVM science program ($<1-20$\,pc) is only possible thanks to the very nearby nature of our targets. 

Key to understanding the LVM-I system is the fact that optical fibers allow for a very narrow range of input f-ratio to minimize focal ratio degradation, and that we are interested in observing emission that is extended on scales that are larger than the angular size of the IFU spaxels. The sensitivity to extended emission of the instrument scales as the etendue, $A\Omega$, of the telescope, with $A$ being the telescope aperture and $\Omega$ the telescope field-of-view. For a fixed f-ratio (as required by the fibers) $\Omega$ scales as $A^{-1}$, and therefore the etendue and the surface brightness sensitivity of the system is independent of the telescope aperture. Put simply, for an IFU with a fixed number of spaxels, fed at a fixed f-ratio, the gains associated with a larger telescope collecting area are cancelled by the losses associated with a smaller field-of-view and and a smaller spaxel solid angle on-sky. The choice of telescope aperture merely selects the desired plate scale of the system. We therefore set the telescope aperture at a value that gives us the desired {\em physical} resolution at the distance of our targets rather than going for the largest possible aperture as astronomers might be used to. 

A second major design driver for the LVM-I is the fact that our main target, the Milky Way, spans the full sky, requiring unusual solutions to the problem of sky subtraction. Also, isolated spectro-photometric calibration stars of sufficient brightness to dominate the flux in our spaxels (which are much larger than the seeing), have a low on-sky surface density of $\sim1$\,deg$^{-2}$ (see Fig.~\ref{fig:skyfields_stdstars}), which makes it difficult to observe a sufficient number of standards inside the FOV of an individual telescope. These two factors push the design towards having separate dedicated telescopes feeding the same spectrographs. Different telescopes are used to obtain sky spectra at large angular distances from the science target pointings, and to observe spectro-photometric standards beyond the reaches of the science IFU telescope field-of-view.

We choose an existing replicable spectrograph design (DESI) rather than develop our own, motivated by the need to shorten the project time frame to go on sky and to keep the project costs within the available SDSS budget. While we are aware that for the purpose of studying ISM kinematics, a resolution of $R>10,000$ would have been desirable, this would have required either compromising on wavelength range (and hence on the science) or developing custom-made cross-dispersed spectrographs which would have exceeded any realistic budget and timeline given the scale of the survey. We therefore opted for the lower-resolution ($R\sim4,000$) DESI spectrographs, and have put a focus on driving the system design to deliver very good line-spread-function (LSF) stability. This permits a precise characterization of the LSF for individual science frames, which permits pushing the measurement of line broadening further below the instrumental spectral resolution limit \citep{law21}. LSF stability is also key for minimizing sky subtraction residuals. Experience with fiber-fed IFU spectrographs like MaNGA, PPAK, and VIRUS-P has shown that the main driver of LSF variability during observations is flexure, and the focal ratio degradation effects caused by the movement of optical fibers as the telescope points in different directions. We therefore favor a static focal plane design in which  the fibers are always stationary. These high-level design drivers lead us to the following solution.

\subsection{Telescopes}
\begin{figure}
\plotone{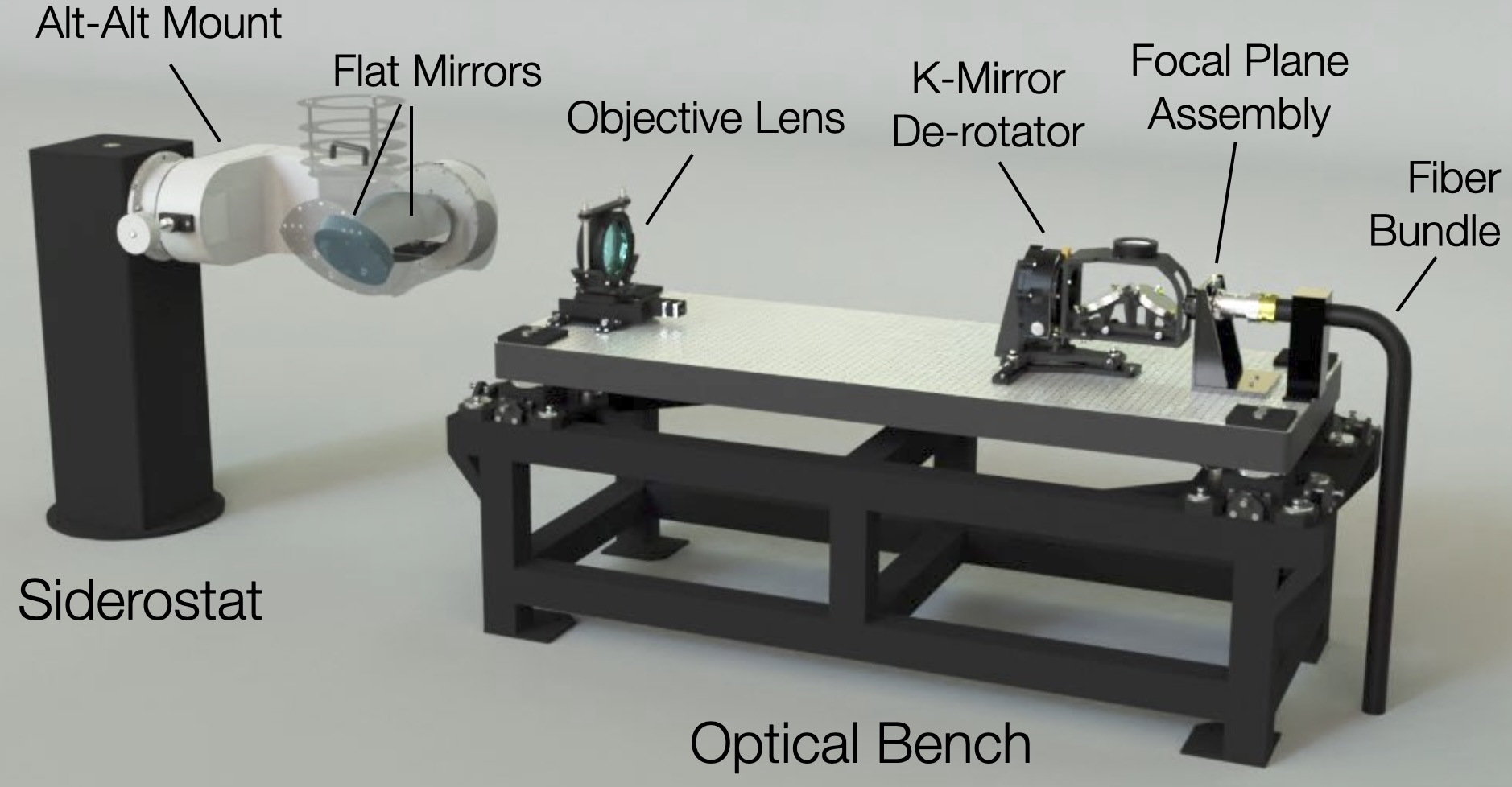}
\caption{An LVM-I telescope and its components. For scale, the optical bench is 210$\times$75\,cm in size}\label{fig:telescope}
\end{figure}

A detailed description of the LVM telescope system is given in \citet{Herbst2024}. LVM-I uses four separate 16.1\,cm aperture refractive telescopes at f/11.42 \citep{Lanz22} delivering a 1.4\degr\ field. The first telescope (called ``Sci'' for ``science'') feeds the main science IFU, a second and a third telescope (``SkyE'' and ``SkyW'' for sky East and West respectively) target sky fields with very faint MW emission, and are used to obtain sky spectra \citep{jones24}, and a fourth telescope (``Spec'' for ``spectro-photometric'') targets bright and very isolated stars used to flux calibrate the spectra and to calculate atmospheric telluric-absorption corrections \citep{kreckel24}.

A major advantage of the LVM is that it is using a stationary fiber system that does not move with the telescopes. In fact, all powered optics in LVM-I are stationary. (see, e.g., \citealp{drory15}, \citealp{law21} and \citealp{bundy22} for discussions of issues of fiber motion and LSF effects in the context of the MaNGA survey).  Our design of choice uses siderostats mounted on commercial PlaneWave L-600 direct drive mounts. Each siderostat (consisting of two flat mirrors) deliver images of the sky to a bench-mounted refractive telescope objective \citep{Lanz22} fitted with K-mirror field de-rotators. The anatomy of an LVM-I telescope is shown in Fig.~\ref{fig:telescope}. All four telescopes are identical in their optical design, with the exception of the calibration telescope which observes individual point sources and therefore does not require a field de-rotator. 

\subsection{IFU Fiber System} \label{sec:instrument-IFU}

\begin{figure}
\plotone{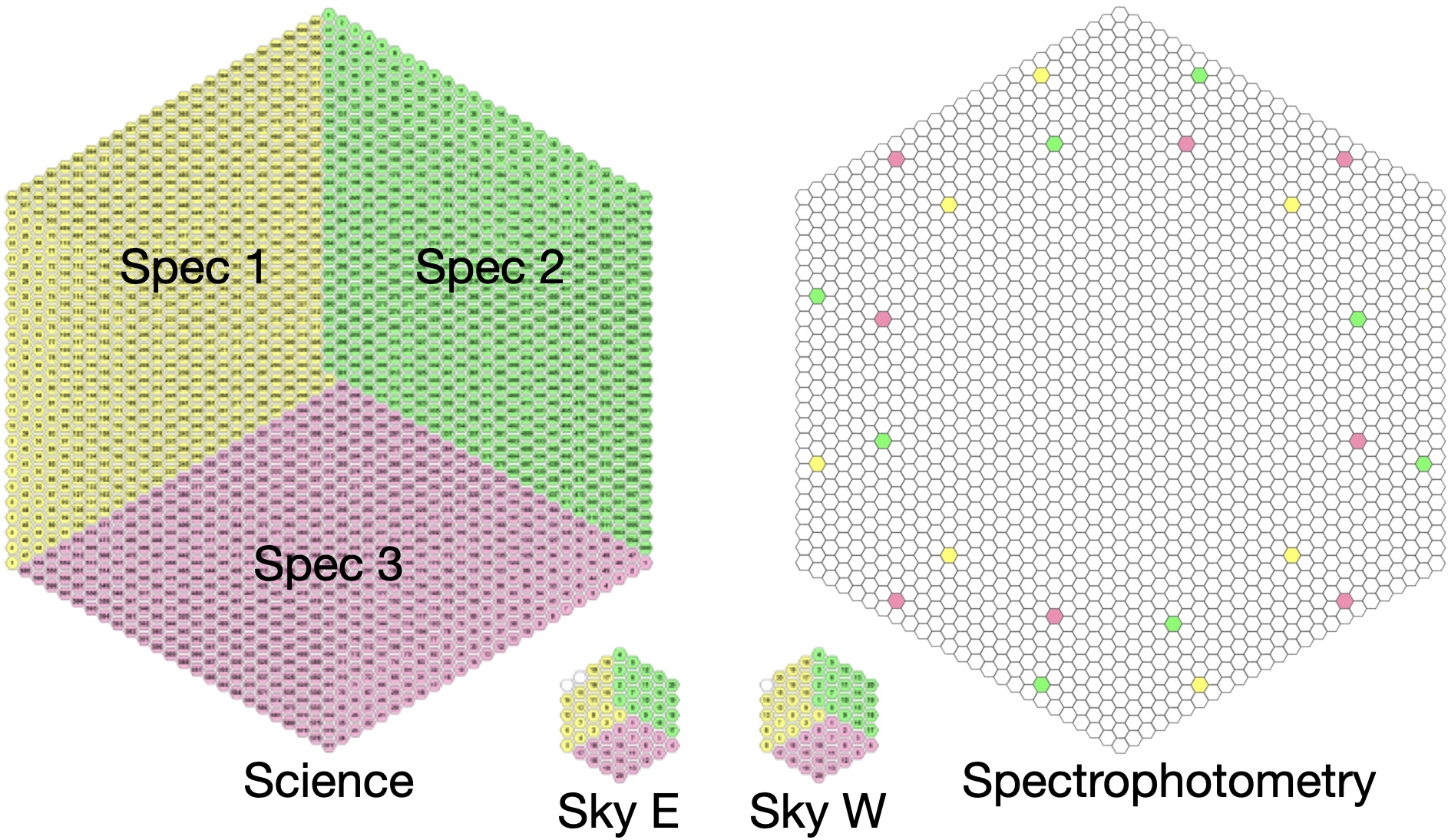}
\caption{The LVM-I integral field units feeding the spectrographs. The science IFU consists of 1801 fibers arranged in 25 hexagonal rings (including the central fiber), and 3 segments feeding one spectrograph each. The two sky telescopes have 59 and 60 fibers respectively and are also divided into 3 segments each feeding the 3 spectrographs. Each block of fibers along the slit is flanked by two a sky fiber on each side, one from each sky telescope. Finally the spectrophotometric telescopes has 24 fibers covered by a rotary shutter allowing us to expose one fiber at a time. Eight fibers go to each spectrograph.}\label{fig:ifus}
\end{figure}

The LVM-I fiber system consists of four fiber-coupled micro-lens array IFUs that are bench-mounted at the fixed focal planes of the telescopes (Figure \ref{fig:ifus}). After leaving the four IFU assemblies the 1944 optical fibers meet in a sorting box (the ``sorting hat'') where they are reorganized into three separate output bundles \citep{Feger20}. Each of these bundles carries 648 fibers consisting of a mix of science, sky, and spectro-photometric calibration fibers from different telescopes, which terminate in a pseudo-slit unit that feeds their light into each of the three LVM-I spectrographs. The fibers are 107\,$\mu$m core Molex FBP fed at f/3.7. All fibers and powered optics are stationary, which together with the temperature stabilized spectrograph chamber (see below) contributes to minimizing LSF variablity. 

The science IFU consists of 1801 fibers fed by a 25-ring hexagonal micro-lens array (see Figure \ref{fig:ifus}) that images the f/11.42 telescope beam down to f/3.7. To prevent diffraction effects at the edge of the hexagonal lenslets, a chrome mask is used to produce an array of circular 35.3\arcsec\ diameter apertures that follows the hexagonal tiling pattern of the micro-lens-array, which has a 37\arcsec\ pitch. This translates into an 83\% fill factor after the mask is installed. The field-of-view of the science IFU is a hexagon with a 30.2\arcmin\ outer diameter, 15.4\arcmin\ long sides, and a total on-sky area of 0.165 deg$^2$. This half a degree IFU FOV pushes the LVM-I into a new regime of ultra-wide-field integral field spectroscopy.

The two sky IFUs have the same optical prescription as the science IFU but their hexagonal micro-lens-arrays only have 5-rings (61 lenslets). The spectro-photometric calibration IFU is identical to the science IFU, but it is sparsely populated with only 24 fibers, distributed along two rings of spaxels (12 in the 19th ring and 12 in the 24th ring, as seen in Fig.~\ref{fig:ifus}). These individual calibration fibers are exposed by actuating a rotating shutter in front of the IFU, allowing the LVM-I to observe typically 12 bright spectro-photometric standard stars around the direction of each science field during an exposure.

Each telescope is fitted with a pair of CMOS acquisition and guiding (AG) cameras of $26.8\arcmin\times 18.4\arcmin$ FOV \citep{2022SPIE12184E..6UH}, fed by two pick-off mirrors flanking the central IFU (except for the Spec telescope which is only fitted with one camera).

\subsection{Spectrographs}

The LVM-I images the fiber spectra using three spectrographs which are identical to the Dark Energy Spectroscopic Instrument spectrographs \citep{10.1117/12.2311996,desi22} except for certain modifications that are described in detail by \cite{konidaris20}. The main difference between the LVM-I and the DESI units is that in our system the slits are denser, carrying 648 fibers instead of 500, and that we adopt a different design for the detectors and dewars. Unlike DESI, the LVM-I detectors are LN2 cooled and are also designed and fabricated by a different vendor.

The spectrographs have a nested dichroic design that splits the light of the fiber spectra into three wavelength channels: blue (b: 3600-5800\,\AA), red (r: 5750-7570\,\AA), and infrared (z: 7520-9800\,\AA). Along each channel volume-phase holographic (VPH) gratings are used to disperse the light and image the fiber spectra through a fast f/1.7 cameras onto 4k$\times$4k CCDs with 15$\mu$m pixels. The detectors are STA~4850 devices, with a 30\,$\mu$m epitaxial variant used for b and r channels, and the 100$\,\mu$m deep depletion variant used for the z channel. We employ Archon controllers for readout \citep{Archon14}. The dewars are custom designs manufactured by Universal Cryogenics, Inc. The CCD/dewar system is contributed by ITL, University of Arizona. The average spectral resolutions across the b, r, and z channels is $R=2700$, 4000, and 4600 respectively.


The spectrographs are housed in a temperature stabilized insulated room with an HVAC system keeping the ambient temperature at $20\pm0.1\degr$C. As mentioned above the detectors are cooled by LN2, which is fed by an automatic filling system connected to an external storage tank.
\section{The LVM Survey} \label{sec:lvmsurvey}

\begin{figure*}
\plotone{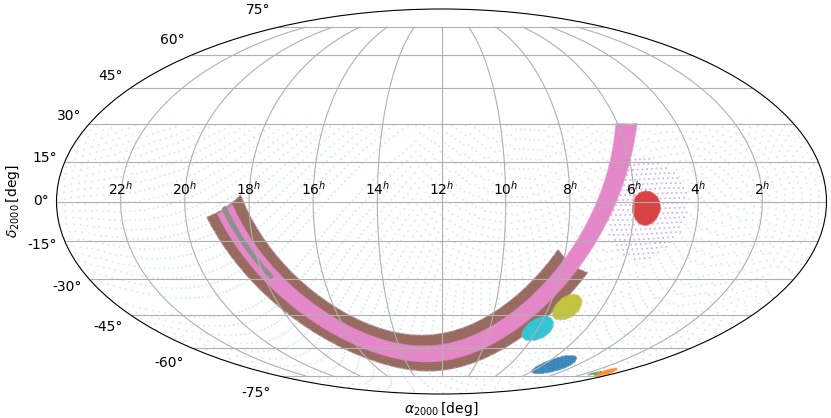}
\caption{LVM baseline 4-year survey footprint and target definition. We divide the MW midplane survey into a high-priority 8$\degr$~wide band (pink), surrounded by a lower-priority 18$\degr$~wide band (brown). We highlight the region observed by THOR near the MW center (gray). In addition, we target a densely sampled region around the Orion nebula (near RA 6h, just south of the plane; red), surrounded by a wider, more sparsely sampled region to cover the entire ionized nebula and its interface to the surrounding ISM. We extend the midplane survey around RA of 8h to cover a wider area in the Gum nebula (cyan, green). We augment the MW disk coverage with a sparse all-sky grid of pointings to sample the out-of-plane targeted when no other target is observable (gray). Finally, we target the LMC (blue) and SMC (orange).}\label{fig:lvm-footprint}
\end{figure*}

\begin{table*}
    \centering
    \begin{tabular}{ccccccl}
    \hline
       Target & Area & \# Spectra & Spaxel & 5$\sigma$~depth & Geometry &\\
              & sq.\,deg. &  & pc & erg\,s$^{-1}$\,cm$^{-2}$\,arcsec$^{-2}$ & & \\
        \hline\hline
        MW mid-plane    & 2700    & 35M & 0.1-1 & $6\times10^{-18}$ (\Halpha)& Center (l,b) = (282\degr,0\degr); (w,h) = (218\degr,8\degr) &\\
        MW plane extension   & 1744    & 22M & 0.1-1 & same & Center (l,b) = (315\degr,0\degr); (w,h) = (150\degr,18\degr) &\\
        Orion       & 132.7   & 1.7M & 0.07 & same & 
        Circle (l,b) = (206.42\degr, -17.74\degr), $r=6.5$\degr&\\
        & & & & & 1/5 fill $r=$19.5\degr&\\
        Gum 1 & 95 & 1.1M & 0.08 & same & Circle RA 109.53\degr, DEC -41.54\degr, $r=5.5\degr$&\\
        Gum 2 & 95 & 1.1M & 0.08 & same & Circle RA 119.75\degr, DEC -50.90\degr, $r=5.5\degr$&\\
        \hline
        MW high latitude & $\dots$ & 1M & ... & $6\times10^{-18}$ (\Halpha) & Sparse grid of single pointings &\\
        \hline
        LMC         & 78.5    & 1M & 10 & $2\times10^{-18}$ (\Halpha; 23.0 $V_{\mathrm{AB}}$) & 
        Circle RA 79.51\degr, DEC -68.53\degr, $r=5\degr$  \\
        SMC         & 16.4    & 200k & 10 & same & 
        Ellipse (RA 13.16$\degr$, DEC -72.80$\degr$, $a=$2.8$\degr$, $b=$1.6$\degr$, $pa=$45\degr)&\\
        & & & & & $\cup$ (RA 20.00$\degr$, DEC -73.20$\degr$, $a=$1.5$\degr$, $b=$1.0$\degr$, $pa=$155$\degr$)&\\
        \hline
    \end{tabular}
    \caption{Overview of LVM targets, coverage, depth and geometry.}
    \label{tab:targets}
\end{table*}

The LVM survey observes the Milky Way and the Magellanic System. In addition, when these targets are not observable we will target a sample of nearby, large apparent size galaxies in the southern hemisphere. In this section, we provide an overview of the baseline 4-year survey extent, targeting, and survey execution and observing strategy.

\subsection{The Milky Way Survey}

We divide the MW midplane survey into a high-priority 8$\degr$~wide band centered on $b=0$, surrounded by a lower-priority 18$\degr$~wide extension. Within the MW, the survey will prioritize survey tiles covering known \HII\ regions to facilitate early science. However, survey simulations predict near complete coverage of the survey area within the planned survey duration.  

The Orion nebula and the Gum nebula are two of the closest and best studied nebulae in the Milky Way. Due to their proximity, they extend far from the disk midplane, and we therefore target them individually. We add a circular region centered on RA 206.42$\degr$ and DEC -17.74$\degr$ with a radius of 6.5$\degr$ to densely sample the core of the Orion nebula, and a larger region of radius 19.5$\degr$\ sampled with a fill factor of 1/5 
to cover the entire ionized nebula and its interface to the surrounding ISM. The Orion region is of high priority, and within the region we observe center-out. Similarly to the core of Orion, we observe the shell and the south-western limb of the Gum nebula. We add two densely sampled circular regions of radius 5.5$\degr$ centered on 
RA 119.75$\degr$, DEC -50.90$\degr$, and RA 109.53$\degr$, DEC -41.54$\degr$, respectively.

All observations in the MW have a nominal exposure time of 900\,s, with the goal of reaching a line sensitivity of $5\sigma=6\times10^{-18}$\,erg\,s$^{-1}$\,cm$^{-2}$\,arcsec$^{-2}$ at 6563\,\AA. The Milky Way is observed at any lunation, with observing constraints of distance to the Moon larger than 60$\degr$, airmass below 1.75, and Earth shadow height\footnote{The Earth shadow height is defined as the distance from the surface of the Earth to the location at which the observational line of sight intersects the Earth's shadow. It is the primary parameter that defines the column of upper atmosphere gas that is directly exposed to solar radiation along the line of sight, and is therefore strongly correlated with the intensity of Geocoronal emission.} of $>1000$\,km to lower the impact of geocoronal \Halpha\ on the MW signal. Due to the relative brightness of \Halpha\ in the Orion and Gum nebulae, the shadow height constraint is relaxed to $>500$\,km, with all other parameters the same as for the rest of the Galaxy. For the fiducial exposure time a small number of survey tiles show saturation in the strong nebular lines. We identify these cases and perform repeat short 10\,s exposures on them to recover the spectra of the bright saturated features.

Finally, we target a sparse grid of all-sky pointings with a 4 degrees spacing through the out-of-disk ISM to be observed when other targets are not observable given the scheduling constraints. Table~\ref{tab:targets} lists the survey geometry and Fig.~\ref{fig:lvm-footprint} shows the survey area on the sky.

\subsection{The Magellanic System Survey}

\begin{figure*}
\plottwo{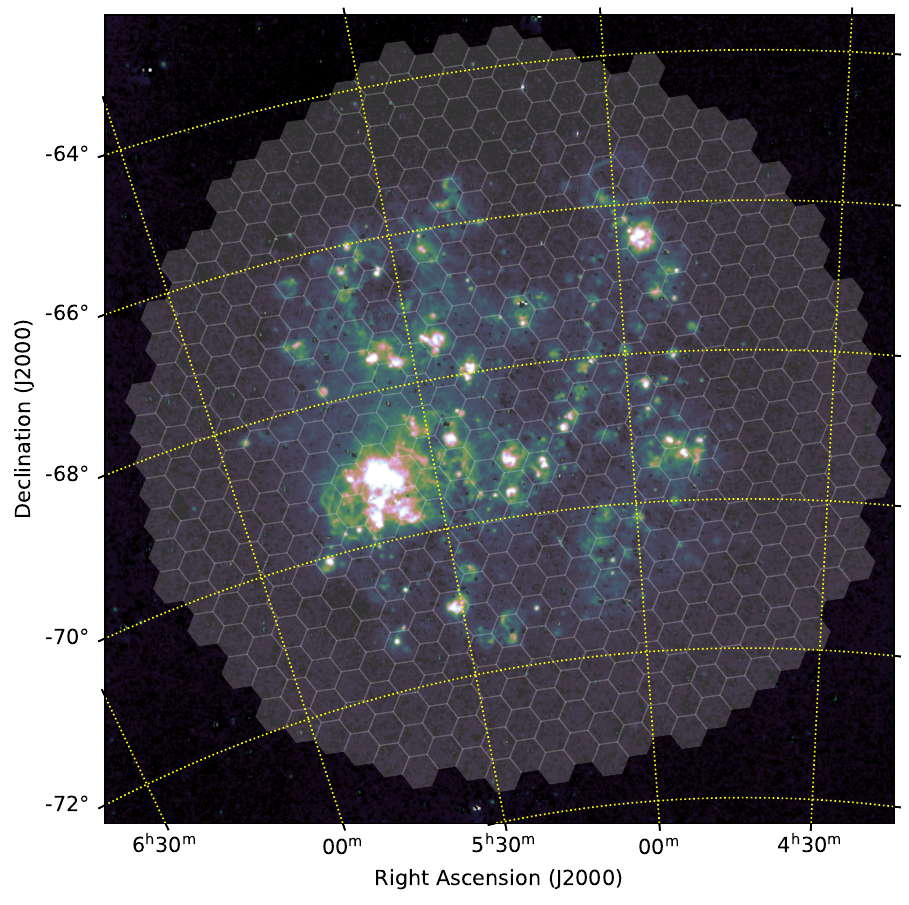}{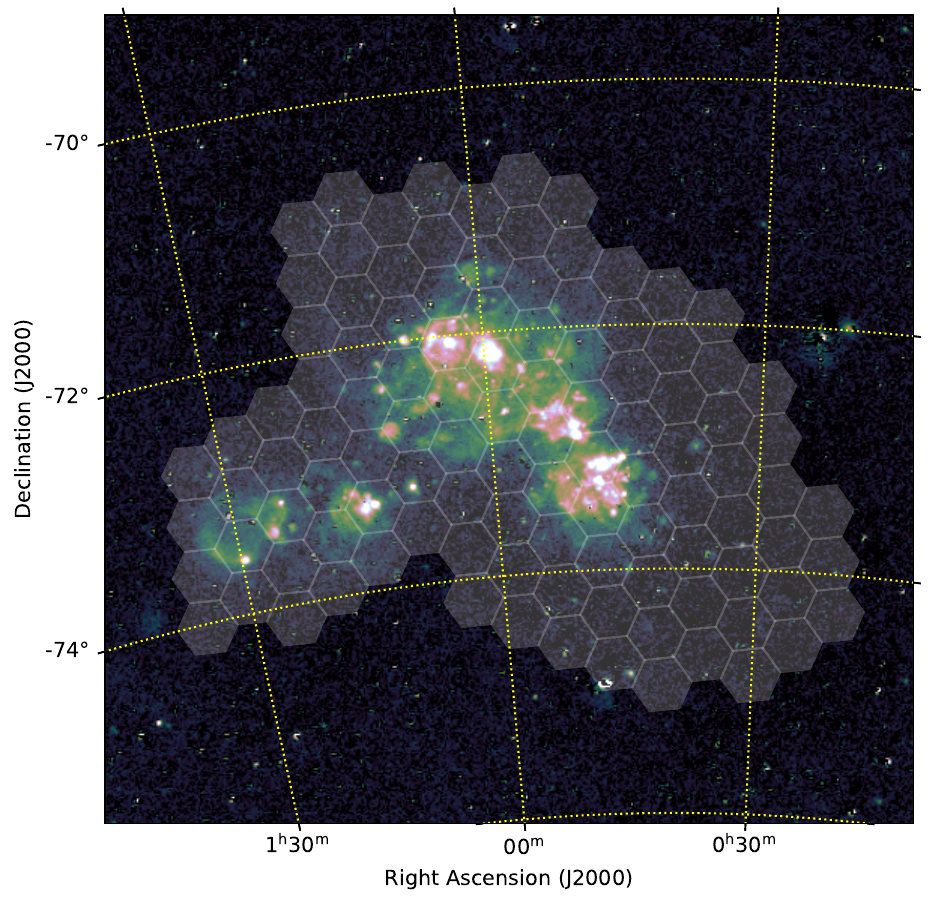}
\caption{LVM Survey area with tile outlines in the LMC (left panel) and the SMC (right panel) overlaid on top of \Halpha\ imaging SHASSA \citep{gaustad01}.}\label{fig:lmcsmc}
\end{figure*}

All observations in the Magellanic System have a nominal exposure time of 8100\,s broken up into 9 exposures of 900\,s each. The exposures are dithered on a 9 point grid similar to the one employed in MaNGA (see \citealp{law16}).
The Magellanic System observations aim to reach an emission line depth of $5\sigma=2\times10^{-18}$\,erg\,s$^{-1}$\,cm$^{-2}$\,arcsec$^{-2}$ at 6563\,\AA, and a continuum surface brightness depth at $5\sigma$ of 23.0\,AB\,mag\,arcsec$^{-2}$ in the V band.

Unlike the Milky Way survey, we restrict the Magellanic System observations to dark/gray time, with a lunation constraint of $<0.25$. We relax the airmass constraint to $\sec z < 2.0$ and observe at distance from the Moon of 45$\degr$\ or more. Because of the velocity offset of the Magellanic Clouds relative to the Earth, geocoronal emission is less of a concern and we do not impose a shadow height limit.

The LMC surveys covers the galaxy to roughly its optical radius, $D_{25}$. We survey a circular region of radius 5$\degr$ centered on RA 79.51$\degr$, DEC -68.53$\degr$.  The SMC survey area is comprised of the union of two ellipses to capture the main part of the SMC disk and the star-forming complexes on the south-eastern side of the galaxy. The main galaxy area is defined as RA 13.16$\degr$, DEC -72.80$\degr$, with semi-major axes $a$ 2.8$\degr$, $b$ 1.6$\degr$ and position angle PA of 45$\degr$. The south-eastern extension is defined in terms of RA 20.00$\degr$, DEC -73.20$\degr$, with semi-major axes $a$ 1.5$\degr$, $b$ 1.0$\degr$ and position angle PA of 155$\degr$. Fig.~\ref{fig:lmcsmc} shows an overlay of the survey area in the Magellanic system on top of \Halpha\ images and Table~\ref{tab:targets} lists the survey geometry.

\subsection{The Nearby Galaxies Survey}
The remarkably large field of view of the LVM is designed to efficiently observe our core survey targets, however it is naturally well suited for other extended nearby galaxies as well. Moving to larger distances, the 35.3\arcsec\ LVM fiber size samples increasingly larger physical scales. At distances of $\sim$20\,Mpc  this corresponds to 3-4\,kpc scales, equivalent to what has been achieved for more distant galaxies in surveys like MaNGA \citep{bundy15} and SAMI \citep{2021MNRAS.505..991C}. 

LVM enables the unique opportunity to resolve complete nearby galactic systems out to $\sim$20 Mpc distances, often in a single LVM footprint. 
Modern IFU instruments remain inefficient when mapping the large apparent sizes ($R_{25}$$\sim$5-10\,arcmin) of such nearby galaxies, with the most ambitious programs (e.g., PHANGS-MUSE; \citealt{2022A&A...659A.191E}) typically covering out to only 0.5\,$R_{25}$.  With LVM we are able to spectroscopically map full galaxy disks,  with the unique advantage of having multi-wavelength data at matched (or significantly better) resolution easily achievable, and often already available in the literature. With ancillary 1\,\arcsec$\sim$100\,pc resolution observations able to isolate individual star-forming regions and ionizing sources, LVM observations will constrain the kpc-scale ionized gas and stellar populations. This enables us to link ancillary observations at small scales with bulk energy injection on large scales, placing direct constraints on how these physical processes relate to global gas inflow and outflow processes. 
These targets are also ideally suited for matched-scale multi-phase observations, linking ionized, atomic and molecular gas properties. 
All targets will be observed with dithering to achieve improved image fidelity and recover the complete flux. 
The addition of these targets to the survey plan also makes effective use of unused telescope time, when our higher priority core science targets are unavailable. 

We have compiled a master list of nearly 500 galaxies within 20\,Mpc that are well suited for LVM observations, and will be observed on a best-effort basis over the course of the survey \citep{kreckel27}. Full completion of this target list is not guaranteed, but survey simulations suggest near complete coverage is feasible. Due to the incompleteness of galaxy catalogs out to 20\,Mpc distances, and the unique opportunities provided by LVM to observe the closest star-forming dwarf galaxies, we have developed our target list by combining two strategies. The first is tailored to the Local Group, and the second focuses on massive star-forming galaxies out to $\sim$20 Mpc.

\subsubsection{Local Group}
\begin{table}[]
    \centering
    \begin{tabular}{l|r r c c}
\hline
Name & RA  & Dec  & Dist  & \# LVM  \\
  &  [hms] &  [dms] &  [Mpc] &  pointings \\
\hline
\hline
WLM  &  0:01:58.1 & -15:27:40  &  1.0 &  1 \\
NGC~55  &  0:14:53.6 & -39:11:48  &  2.1 &  7 \\
NGC~247  &  0:47:08.3 & -20:45:36  &  3.6 &  7 \\
NGC~253  &  0:47:34.3 & -25:17:32  &  3.9 &  7 \\
NGC~300  &  0:54:53.5 & -37:40:57  &  2.1 &  7 \\
IC~1613  &  1:04:47.8 & +2:08:00  &  0.7 &  1 \\
M33  &  1:33:50.9 & +30:39:37  &  0.7 &  24 \\
NGC~625  &  1:35:05.0 & -41:26:11  &  3.9 &  1 \\
NGC~1313  &  3:18:16.1 & -66:29:54  &  4.1 &  7 \\
NGC~2915  &  9:26:11.5 & -76:37:35  &  3.8 &  1 \\
Sextans~B  &  10:00:00.1 & +5:19:56  &  1.4 &  1 \\
NGC~3109  &  10:03:07.2 & -26:09:36  &  1.3 &  1 \\
UGC~05456  &  10:07:19.7 & +10:21:44  &  5.6 &  1 \\
Sextans~A  &  10:11:00.8 & -4:41:34  &  1.3 &  1 \\
IC~3104  &  12:18:46.1 & -79:43:34  &  2.3 &  1 \\
GR~8  &  12:58:40.4 & +14:13:03  &  2.1 &  1 \\
NGC~4945  &  13:05:26.1 & -49:28:16  &  3.8 &  7 \\
NGC~5102  &  13:21:57.8 & -36:37:47  &  3.4 &  1 \\
NGC~5128~(CenA)  &  13:25:28.9 & -43:01:00  &  3.8 &  7 \\
ESO~324-024  &  13:27:37.4 & -41:28:50  &  3.7 &  1 \\
NGC~5253  &  13:39:55.8 & -31:38:24  &  3.6 &  1 \\
ESO~325-011  &  13:45:00.8 & -41:51:32  &  3.4 &  1 \\
ESO~383-087  &  13:49:18.8 & -36:03:41  &  3.5 &  1 \\
Circinus  &  14:13:09.3 & -65:20:21  &  4.2 &  1 \\
ESO~274-001  &  15:14:13.5 & -46:48:45  &  3.1 &  1 \\
IC~4662  &  17:47:06.3 & -64:38:25  &  2.4 &  1 \\
NGC~6822  &  19:44:57.7 & -14:48:11  &  0.5 &  1 \\
IC~5152  &  22:02:41.9 & -51:17:43  &  2.0 &  1 \\
NGC~7793  &  23:57:49.4 & -32:35:24  &  3.9 &  1 \\
\hline
    \end{tabular}
    \caption{Local Group targets (parameters taken from \citealt{Karachentsev2013})}
    \label{tab:LG}
\end{table}

This sample of 29 galaxies (Table \ref{tab:LG}) is drawn from existing highly complete catalogs of local galaxies \citep{Karachentsev2013,Kennicutt2008} and defined with the following cuts on global galaxy properties: 

\begin{enumerate}
\item Observable from the south. An upper declination limit of +15 degrees ensures that galaxies are observable at an airmass $<$ 1.5 for $\sim$3h. We make an exception for the case of M33.
\item High physical resolution. We select all galaxies that have reported distance of less than 4\,Mpc in at least one of the two catalogs, plus a few additional well-studied targets that are close to this limit (NGC~1313, UGC~05456, and Circinus).  This allows us to connect ionizing sources to the $\sim$kpc ISM properties. For reference, at this distance limit GALEX imaging corresponds to 5\arcsec$\sim$100\,pc, ground-based optical imaging corresponds to 1\arcsec$\sim$20\,pc, and individual LVM fibers correspond to 35.5\arcsec$\sim$700\,pc (i.e.\ CALIFA-like resolution).
\item Strong likelihood for emission line detections. From \cite{Kennicutt2008}, we require a total integrated H$\alpha$ + [NII] flux above 10$^{-13}$ erg/s/cm$^2$.  
\end{enumerate}

This sample is not cut on stellar mass, and the resulting stellar masses range from 10$^{6.8}$-10$^{11}$ M$_\odot$ (assuming a K-band mass-to-light ratio of 1). Most galaxies are covered out to R25 with a single LVM footprint.  M33 requires more extensive coverage with 24 LVM pointings, and seven additional galaxies (NGC~55, NGC~247, NGC~253, NGC~300, NGC~1313, NGC~4945, NGC~5128/CenA) are covered with seven pointings. 

\subsubsection{Local 20 Mpc sample}

This sample of nearly 400 galaxies is drawn principally from \cite{Karachentsev2013} and \cite{Leroy2019}.  Galaxies are selected to be observable (declination $\lessapprox$ 15$^\circ$), massive (M$_* \gtrapprox 10^{9}$ M$_\odot$), star-forming (SFR $>$ 0.1 M$_\odot$/year; when estimates are available), and nearby (D $<$20 Mpc). 
These limits are not strict, particularly as large distance uncertainties are common, and alternate literature sources are taken into account for some individual targets. Single LVM pointings are appropriate for all galaxies, providing full coverage of the optical disk out to R25.  The full sample table will be included in a future publication that provides complete details of the final LVM Nearby Galaxy survey \citep{kreckel27}.   

\subsection{Sky Fields}

\begin{figure*}
\plottwo{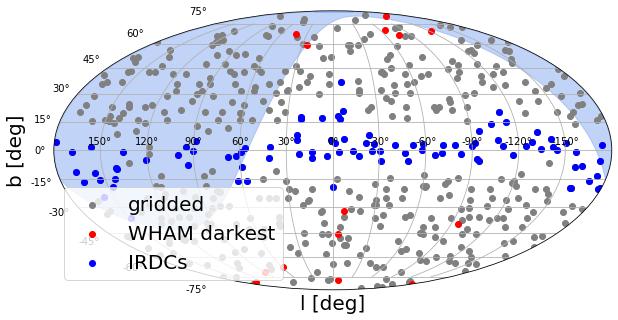}{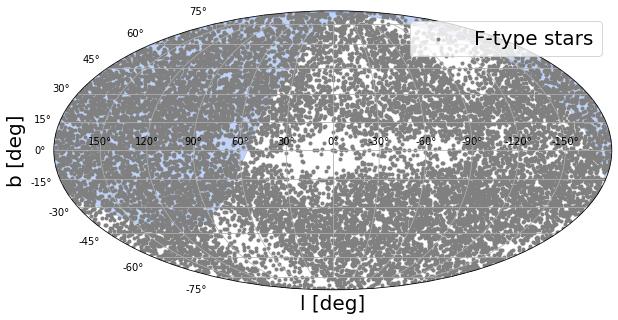}
\caption{Candidate sky fields (left) and spectrophotometric calibration stars (right). Areas of the sky not observable by LVM at LCO are shaded in blue. Left: Grey dots form an all-sky grid of sky fields. Red dots are the WHAM darkest regions with \Halpha\ flux below 0.5\,R, blue dots are the IRDC fields (see text for explanation of the selection). 
Right: The location of bright, isolated F-type stars (grey dots) are selected from Gaia DR3.  We can find ensembles of 12 calibration stars within 5-8 degrees of any science field.
}\label{fig:skyfields_stdstars}
\end{figure*}


The Milky Way fills the sky. Subtracting the night sky spectrum requires us to find places in the sky where the Milky Way emission is faint enough to be considered zero at our survey's depth. However, night sky lines are also highly variable spatially, leading us to construct a grid of optimal sky fields for any 5-10 degree position on the sky. Our sky field identification and selection strategy balances both of these concerns. Fig.~\ref{fig:skyfields_stdstars} (left) shows the distribution of our sky field candidates on the sky.

\textit{WHAM darkest sky fields:} We use the WHAM data \citep{haffner03} to identify regions of exceptionally low Milky Way emission, with \Halpha\ below 0.5\,R\footnote{The Rayleigh (R) is a unit of flux ($F$) defined as the integral over all solid angles (4$\pi$) of an intensity (i.e.\ a surface brightness) $\mu=10^6$\,photons\,cm$^{-2}$\,sr$^{-1}$\,s$^{-1}$. So $F$\,[R]=4$\pi\times10^{-6}\mu$\,[photons\,cm$^{-2}$\,s$^{-1}$]. For H$\alpha$ photons with energy $h\nu_{\rm H\alpha}$, 1\,R corresponds to a surface brightness $\mu_{1R}=5.67\times10^{-18}$\,erg\,s$^{-1}$\,cm$^{-2}$\,arcsec$^{-2}$.} (red points in Fig.~\ref{fig:skyfields_stdstars}, left). This corresponds to 1/3rd of the MW survey 5$\sigma$ depth and is comparable to the depth of the Magellanic System survey. There are only a small number (14) of these, to ensure frequent repeat observations for geocoronal monitoring.

\textit{Gridded sky fields: } A healpix\footnote{https://healpix.sourceforge.io/} division of the sky with {\tt nside = 8} results in 768 grid locations with a separation of 7.5\degr\ between grid centers. For each position, we optimize the exact field center within the  healpix to minimize the Milky Way \Halpha\ foreground, as measured from WHAM, and the integrated stellar light, as measured from summing individual Gaia sources. This results in maximum separations of 15\degr\ between grid locations. 
When possible, we preferentially select positions corresponding to Infrared Dark Clouds (IRDCs) closer to the midplane of the disk (blue points in Fig.~\ref{fig:skyfields_stdstars}, left), selected from \citet{dobashi11}. For these IRDCs, we require a large area $> 71$\,arcmin$^2$ (twice the size of the sky IFU), $A_{\rm V} > 5$\,mag, and a separation of at least 5\,arcmin (the sky IFU size) from sources in the WISE \HII\ region catalog \citep{Anderson14}.

Our initial strategy is to always point one sky telescope to one of the WHAM darkest sky fields, as these are critical for the geocoronal monitoring and subsequent subtraction (see detailed discussion of the sky subtraction strategy in \citealp{jones24}). However, these will be at larger separations from the science fields.
In parallel we use the other sky telescope to observe the closest gridded sky field to our science tile. During commissioning and early observations, we validate and calibrate the sky fields, particularly the IRDC fields, against the WHAM darkest regions, and remove fields with high Milky Way foreground emission or unacceptable stellar contamination from the catalog. 

\subsection{Spectrophotometric Calibration Stars}


The LVM spectrophotometric calibration procedure is a direct descendant of the calibration strategy employed for the MaNGA survey, described in detail in \citet{Yan16} which achieved better than 5\% spectrophotometry across the MaNGA wavelength range of 3600-10000\,\AA. As in MaNGA, we rely on F-type stars for spectrophotometry as these are abundant, bright, and have relatively well understood stellar atmospheres \citep{Kurucz1992}. Unlike MaNGA, we now have GAIA DR3 \citep{GAIADR3} available to us, including BP/RP and RV spectra, to select suitable stars.

To refine the selection of F-type stars, we select based on $T_{\mathrm{eff}}$ from low res spectroscopy as well as high-res RV spectroscopy. We select bright stars with $5 < G < 9$ mag, parallax error less than 20\%, no known variability, $6000\,\mathrm{K} < T_{\mathrm{eff}}< 7500\,\mathrm{K}$, and publicly available BP/RP spectra.

Based on this first selection, we further select only those stars which are isolated enough to dominate the flux across the BP/RP spectrum by at least a factor of 100 in our 35\,arcsec diameter spaxel, i.e.\ the sum of the flux of all other stars down to the GAIA limit is less than 1\% of the flux of the bright F-type star.

This final catalog of stars provides us with a sufficient density of bright calibration stars for all science targets (Figure \ref{fig:skyfields_stdstars}, right). A limited number of sources are found directly towards the inner Galaxy mid-plane, as crowding makes it challenging to find isolated stars, aa well as towards the ecliptic poles as they do not yet have BP/RP spectra available due to the Gaia scanning laws and hence small number of visits in DR3 \citep{GAIAPBRP}.  Overall, the median separation between calibration sources in our bright isolated star catalog is a few degrees, and we can find ensembles of 12 calibration stars within 5-8 degrees of any science field (12 stars is a sufficient number to achieve our calibration goals, as demonstrated by \citet{Yan16} and verified during LVM commissioning).

Spectrophotometric calibration uses the GAIA BP/RP (XP) spectra for the overall calibration, and stellar atmosphere fitting for the high-frequency features as well as for telluric absorption correction. A detailed description of the calibration procedure and performance is given by \citet{kreckel24}. More than 90\% of exposures cycle through 12 standard stars, and 100\% of the exposures have at least 10 stars. Given the exposure time of 900\,s, each standard gets exposed for about 50\,s, with the remainder of the time spent slewing and acquiring the stars. Further optimization of the system will allow for faster acquisitions and hence longer exposures for each standard star (see \citealp{sanchez-gallego24}).

\subsection{Survey Execution}

The LVM-I began commissioning in July 2023, and the LVM survey started in Nov 2023. As of the writing of this paper, the survey operations are funded through May 2027. The facility is starting operations with a remote observer present at all times to open and close the enclosure and monitor the weather and the system for any critical issues. Target selection, scheduling, target acquisition, guiding, and cycling through the spectro-photometric standards during an exposure is fully automated and does not require any human intervention. The plan is for the facility to transition to a fully autonomous operation during 2024. An in-depth description of the operations procedures and software are given in \citet{sanchez-gallego24}.
\section{LVM Data} \label{sec:data}

\subsection{Early Data}

To the east, we observe the iconic Horsehead Nebula, which is associated with the  Orion B molecular cloud (L1630) and in the process of being photo-disassociated. At each position in this map we obtain a full optical spectrum, for a total of $\sim$182,000 spectra.
These 101 tiles were observed on fourteen different nights over the course of more than two months, with varying observing conditions. The remarkable uniformity of this mosaic speaks to the high level of precision we are able to achieve in our flux calibration and astrometry.

\begin{figure*}
\plotone{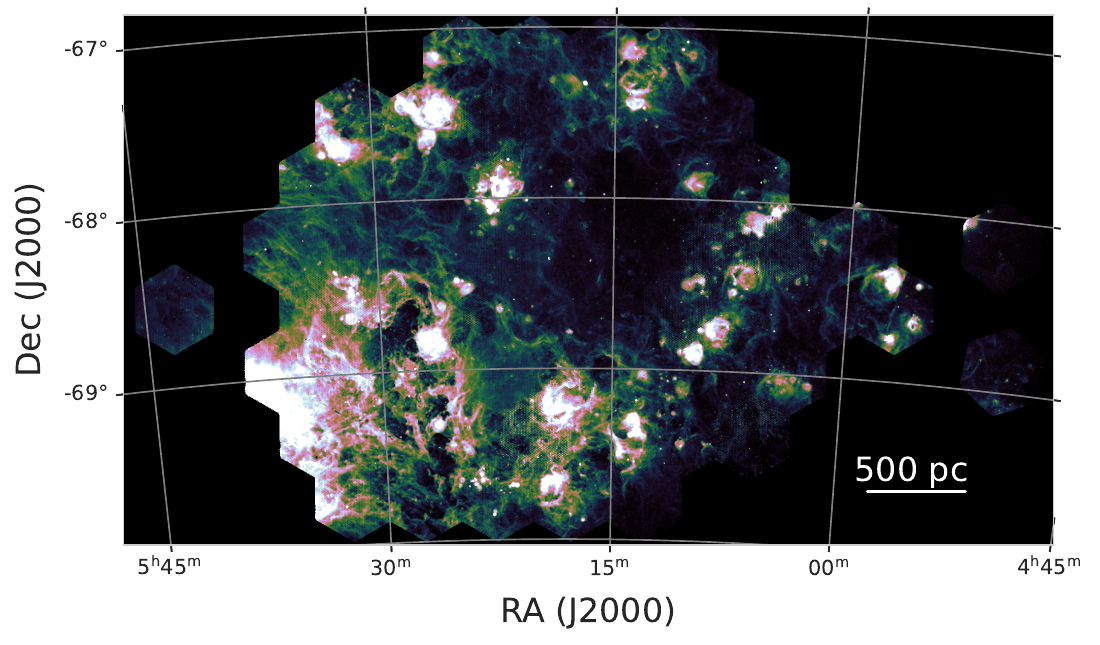}
\caption{58 LVM survey tiles reveal filamentary diffuse structures and bright ionized nebulae in our continuum subtracted H$\alpha$ map, covering $\sim$3\,kpc of the LMC. These maps resolve structures on scales of 10~pc, building towards a panoptic view of this galaxy.}\label{fig:lmc_data}
\end{figure*}

Our early survey operations have also marked the start of our coverage of the LMC, and our survey simulations suggest we will achieve full coverage (see Table \ref{tab:targets}) within the duration of the project. Figure \ref{fig:lmc_data} shows 58 tiles, each observed in nine dither positions,
to create detailed maps of the filamentary nebular emission that threads the ISM of this galaxy. Individual filaments can be identified at 10~pc scales, and are traced in multiple lines (e.g., \oii, H$\beta$,  \nii, H$\alpha$,  \sii).  H$\alpha$ bright photoionized H\textsc{ii} regions can also be identified and resolved, such as the well studied star-forming complex N44 \citep{Oey1995}. Our IFU spectroscopy will provide detailed kinematic constraints on the energetic input from the well studied powering stellar sources \citep{McLeod2019} for a comprehensive sample of star-forming regions across this low metallicity galaxy.  The combination of multiple emission lines, along with fainter auroral line detections,  will also provide detailed abundances of these H\textsc{ii} regions, allowing us to directly map the diffusion and mixing of metals into the disk. In this way, we will build up a comprehensive picture of where energy and material are deposited into the ISM at the energy injection scale across this entire galaxy. 

\begin{figure*}
\plotone{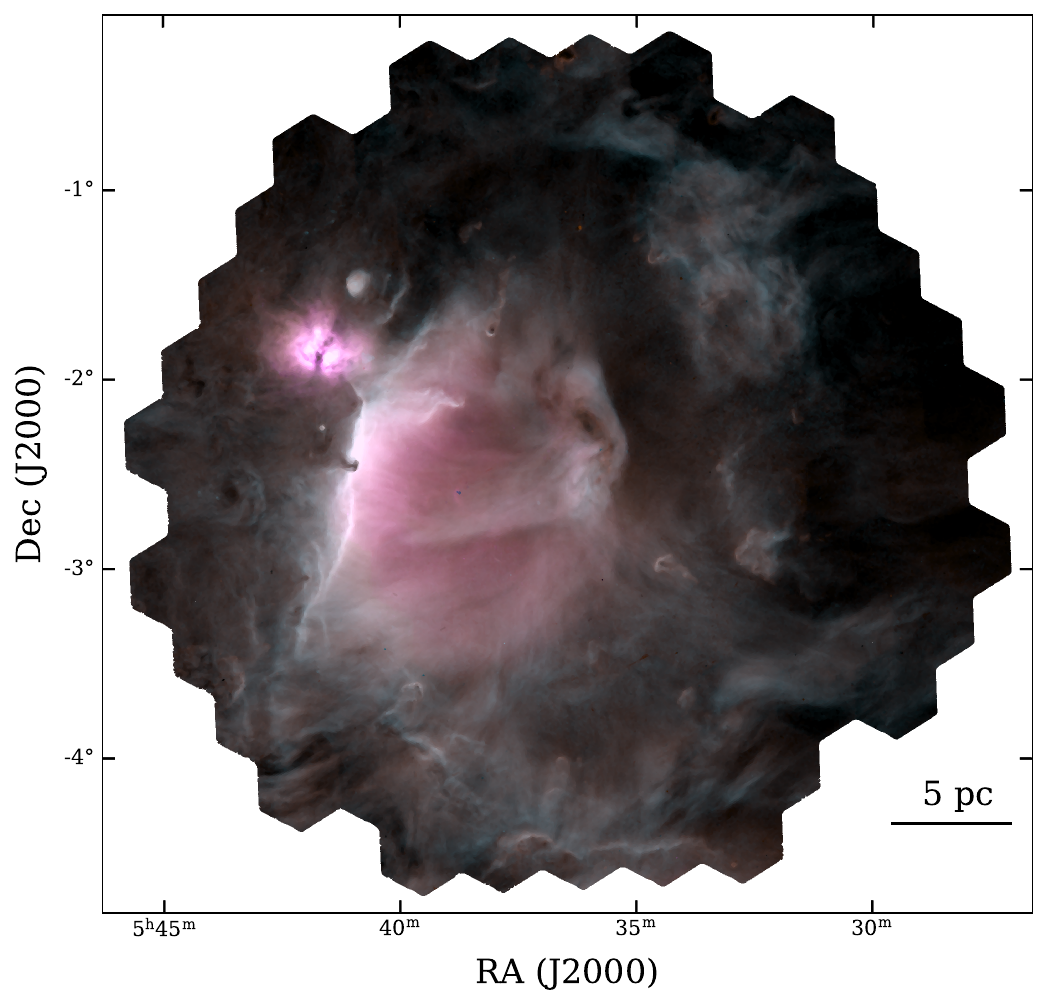}
\caption{Initial 101 LVM survey tiles covering the HII region IC~434 within Orion. Continuum subtracted line emission ([SIII]9532 in magenta, [SII]6717+6731 in blue and H$\alpha$ in orange, tracing high and low ionization zones and average nebula emission, respectively) reveals beautiful ionized gas structures. This map provides $<$0.1 pc scale views, capable of tracing changes in both gas physical conditions (density, temperature, abundances) and kinematics. }\label{fig:orion}
\end{figure*}

\subsection{Pilot Data Analysis Pipeline}

\begin{figure*}
 \minipage{0.99\textwidth}
 \includegraphics[width=18cm,trim={30 85 15 0},clip]{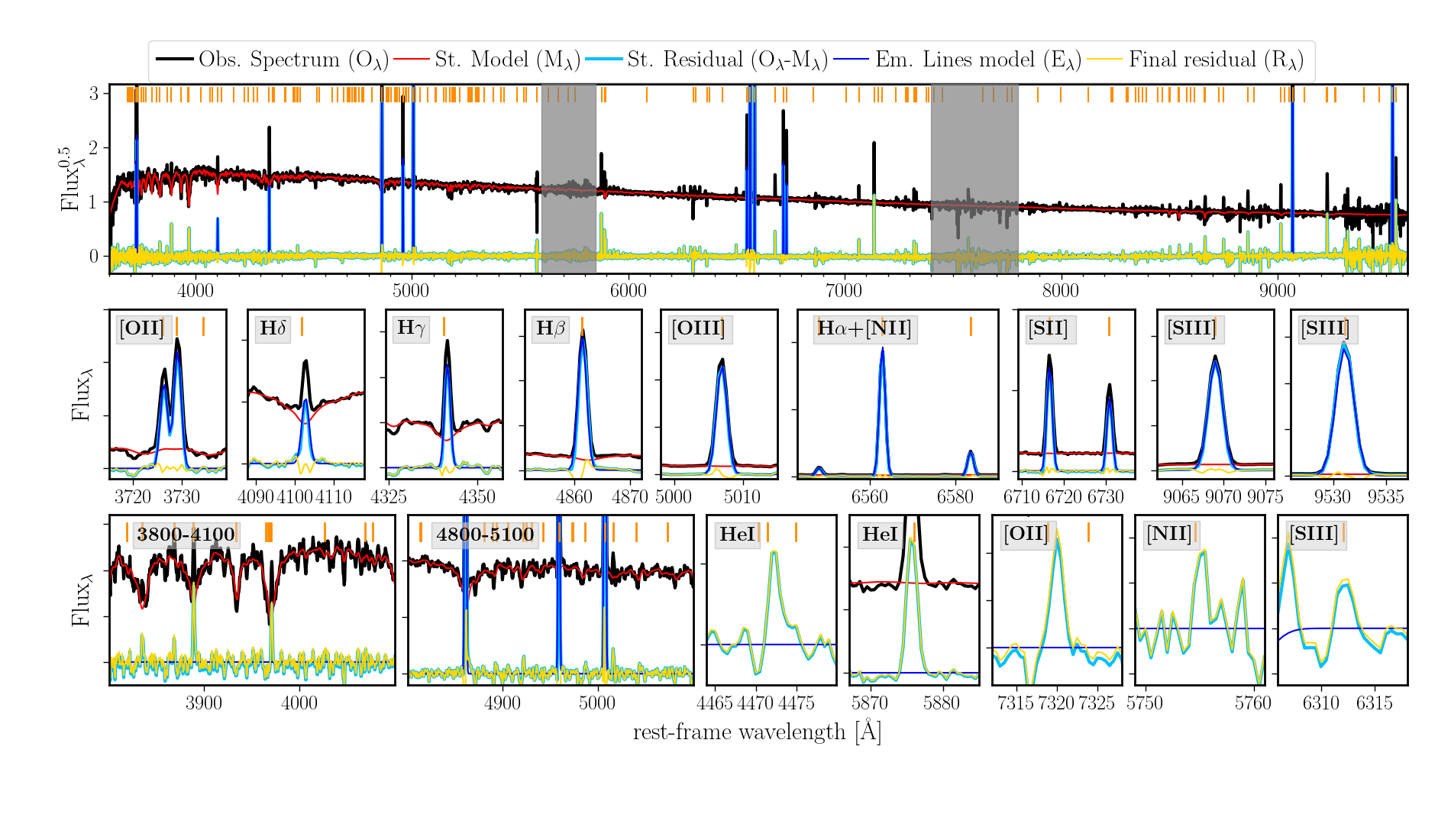}
 \endminipage
\caption{Example of the fitting procedure performed by the pDAP. Top panel shows the integrated spectrum within the central pointing of the Rosette nebula (back line), together with the best-fit stellar template (red line), the residual after subtracting this model (cyan line), the multi-Gaussian model of the ionized gas emission lines (blue line), and the total residual (the result of subtracting the stellar and emission line models from the original spectrum). The small orange vertical lines indicate the emission lines the pDAP is attempting to extract (flux, velocity, velocity dispersion and equivalent-width) using a weighted-moment analysis, as described in the text. We adopt a square-root scaling in the top panel to compress the large dynamic range of the data. The panels in the middle row show zoom-ins around the strongest emission lines, now in linear scaling: [\ion{O}{II}]3726,28 , H$\delta$, H$\gamma$, H$\beta$, [\ion{O}{III}]5007, H$\alpha$+[\ion{N}{II}]6548,83, [\ion{S}{II}]3717,31, [\ion{S}{III}]9069 and [\ion{S}{III}]9531, from left to right. Finally, the bottom row shows a zoom-in around stellar absorption features to illustrate the quality of the template fit: 3800-4100\AA and 4800-5100\AA. The remaining four panels show the detections of weak emission lines, including (i) HeI4471 and HeI6678, two lines used to determine the He abundance \citep[e.g., ][]{valerdi2021}, (ii) [\ion{O}{II}]7318, a line tracing the presence of SNR \citep[e.g.,][]{CF2021}, and (iii) two auroral lines frequently adopted to determine the electron temperature, [\ion{N}{II}]5755 and [\ion{S}{III}]6312 \citep[e.g.,][]{marino13}.}\label{fig:dap_spec}
\end{figure*}

The survey was designed from the very beginning with the goal of providing not only the raw and reduced data, but also high-order data products that allow the community to make use of the observations easily and homogeneously. We follow the path of most recent IFS surveys such as CALIFA \citep{2012A&A...538A...8S}, SAMI \citep{sami}, and in particular MaNGA \citep{bundy15}. All of them have provided high-level data products including observational and/or physical parameters extracted from the observed spectra  \citep{Sanchez2016,LZIFU2016,Westfall2019}. 

Following the Data Reduciton Pipeline (DRP) that produces reduced and calibrated row-stacked spectra, we are developing a data-analysis pipeline (DAP) that decomposes the observed spectra into the stellar (and nebular) continuum, and the ionized gas emission lines, and then extracts a set of observational and physical quantities that characterize each component.

Inside the Local Group, and specially in the MW, individual LVM fibers do not sample scales large enough to contain a representative sample of the local stellar population. In fact, fibers often contain single stars to few stars, or parts of single star clusters. This requires changing the way in which the stellar continuum is typically modeled in IFS observations of nearby galaxies. For the LVM-DAP, we develop a strategy that we name "resolved stellar populations" (RSP), described in \citet{mejia24b}. We do not model the stellar spectra using linear combinations of single stellar populations (SSPs), but a library of individual stellar spectra covering a wide range of stellar properties (T$_e$, log(g), [Fe/H] and [$\alpha$/Fe]).

The current version of the LVM-DAP (pilot DAP) uses the {\sc pyFIT3D} algorithms included in the {\sc pyPipe3D} package \citep{pypipe3d}. A complete description of the pipeline will be presented in \citet{sanchez24}. The pDAP analyzes the LVM-data pointing-by-pointing, using as inputs the spectra and errors provided by the data-reduction pipeline. For each individual spectrum it performs the following steps: 

We first estimate the systemic velocity, v$_\star$, velocity dispersion, $\sigma_\star$ and dust attenuation, A$_\mathrm{V,\star}$, of the stellar component by fitting the continuum with a linear combination of only four stellar spectra that are shifted in velocity, broadened, and dust attenuated, while masking the strongest emission lines.  Once these non-linear parameters are determined, a first model of the stellar component is subtracted, and a set of strong emission lines are fit using Gaussians.
We subtract these emission lines, and then again fit the remaining continuum, this time with a larger stellar template library covering a wide range of stellar properties, yielding our model of the stellar spectrum. This stellar spectrum is then subtracted from the original spectrum. The remaining ``gas-only'' spectrum is then analyzed using a weighted-moment non-parametric procedure to estimate the integrated flux, velocity, velocity dispersion, and EW of the a pre-defined set of 192 emission lines. The parameters derived for the stellar and ionized gas components in each of the previous steps are stored in a set of FITS tables that are packed into a single multi-extension FITS-file and the model spectra are stored in a row-stacked spectra (RSS) format. For the details of the algorithms and their implementation we refer the reader to \citet{sanchez24} and \citet{Lacerda2022} \footnote{http://ifs.astroscu.unam.mx/pyPipe3D/}.

\begin{figure*}
 \minipage{0.99\textwidth}
 \includegraphics[width=6cm,trim={10 55 20 50},clip]{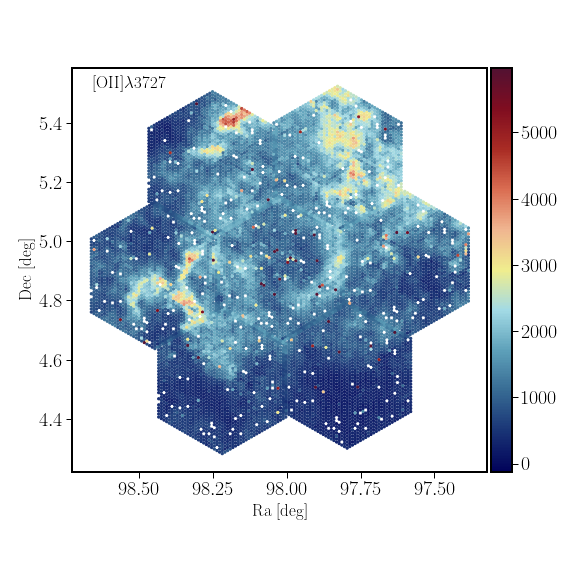}\includegraphics[width=6cm,trim={10 55 20 50},clip]{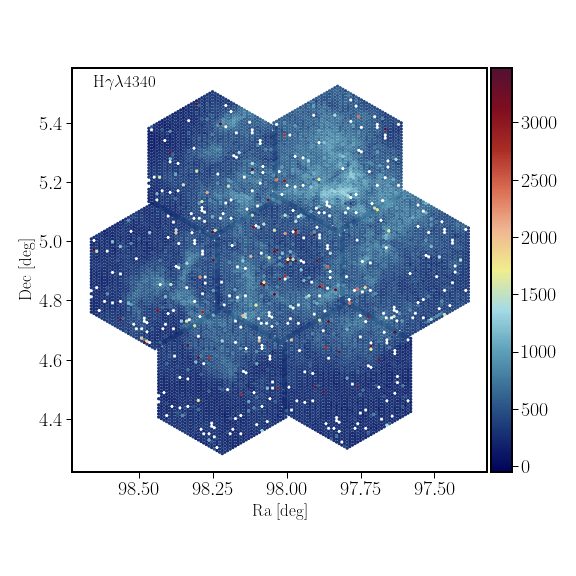}\includegraphics[width=6cm,trim={10 55 20 50},clip]{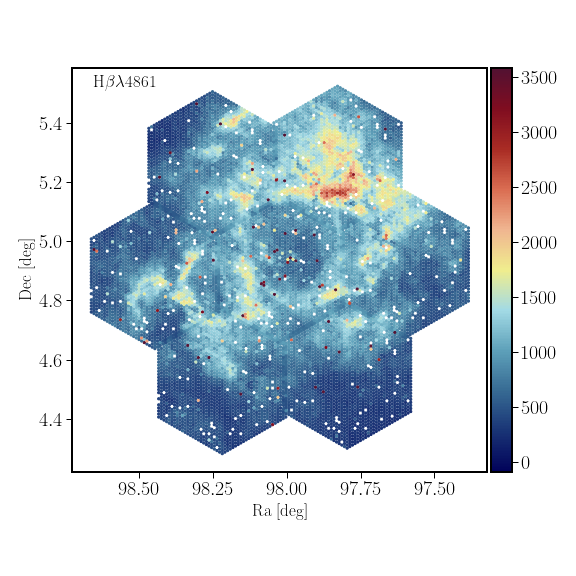}
 \includegraphics[width=6cm,trim={10 55 20 50},clip]{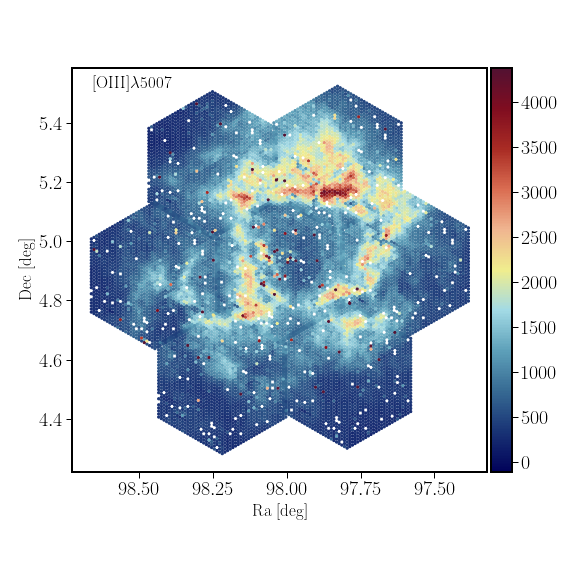}\includegraphics[width=6cm,trim={10 55 20 50},clip]{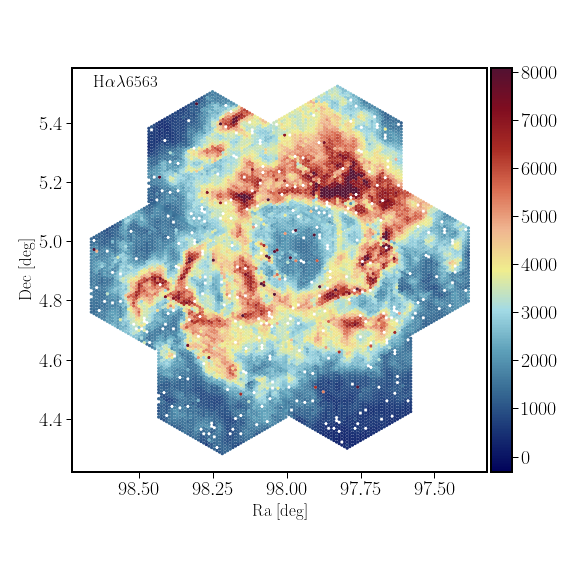}\includegraphics[width=6cm,trim={10 55 20 50},clip]{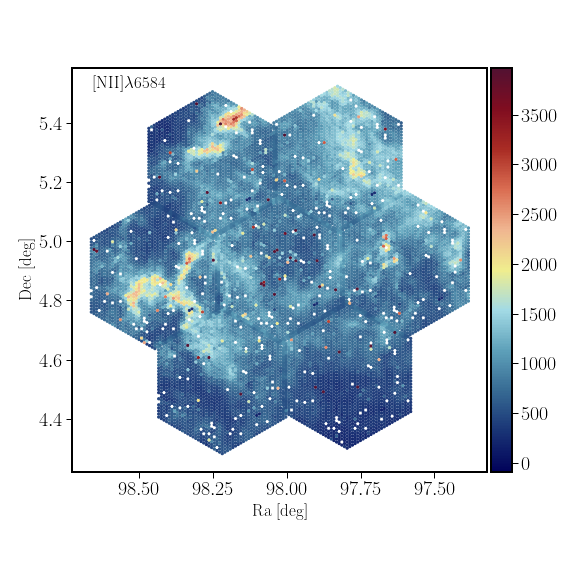}
 \includegraphics[width=6cm,trim={10 55 20 50},clip]{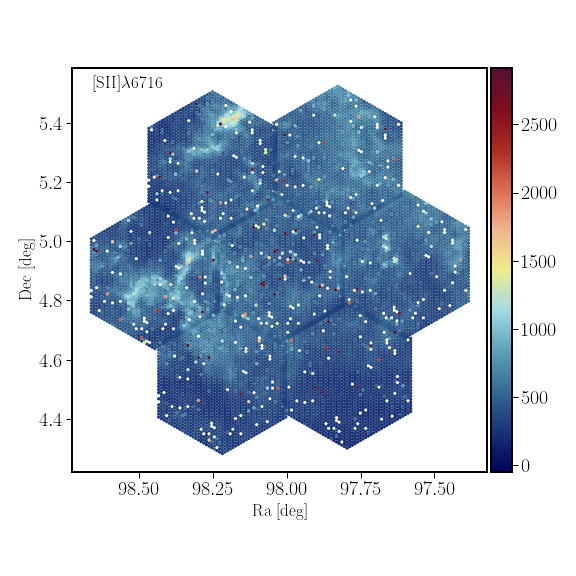}\includegraphics[width=6cm,trim={10 55 20 50},clip]{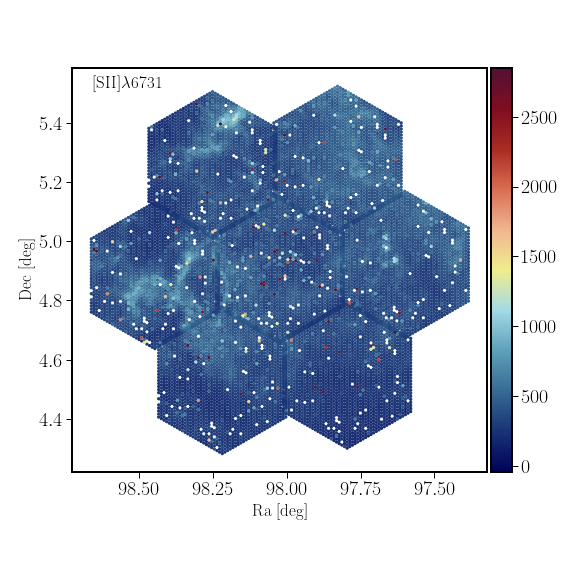}\includegraphics[width=6cm,trim={10 55 20 50},clip]{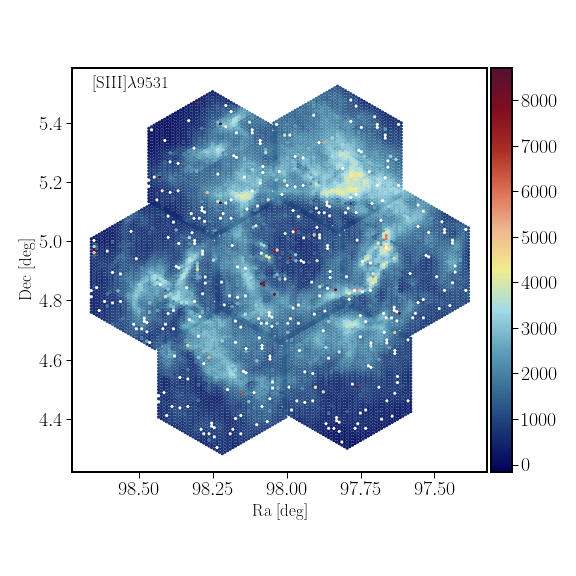}
 \endminipage
\caption{Flux intensity maps in units of \fCAL\ of a set of emission lines derived using the weighted-moments analysis included in the LVM pDAP for the Rosette Nebulae. Each panel comprises a mosaic of 7 different hexagonal pointings, covering an area with a diameter of $\sim$1\degr. The flux extracted for each of the $\sim$1800 fibers (per pointing) is shown as an individual color-coded circles (i.e., no interpolation is performed). The intensity range has been adjusted for each emission line, covering between the mean value minus 10\% and the mean value plus 3$\sigma$.}\label{fig:dap_RT1}
\end{figure*}

\begin{figure*}
 \minipage{0.99\textwidth}
 \includegraphics[width=6cm,trim={10 55 20 50},clip]{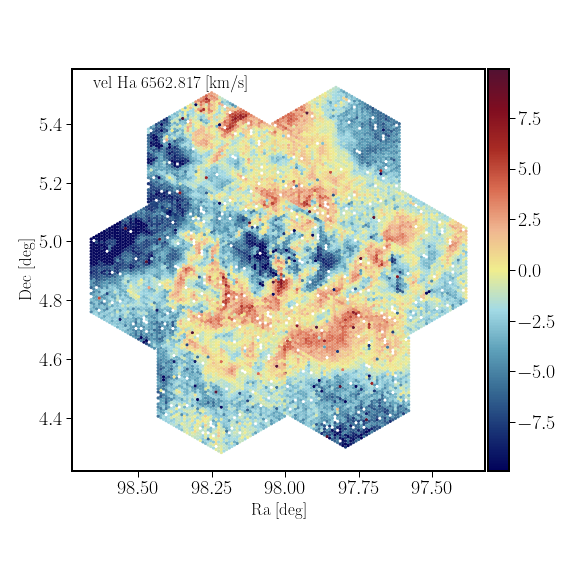}\includegraphics[width=6cm,trim={10 55 20 50},clip]{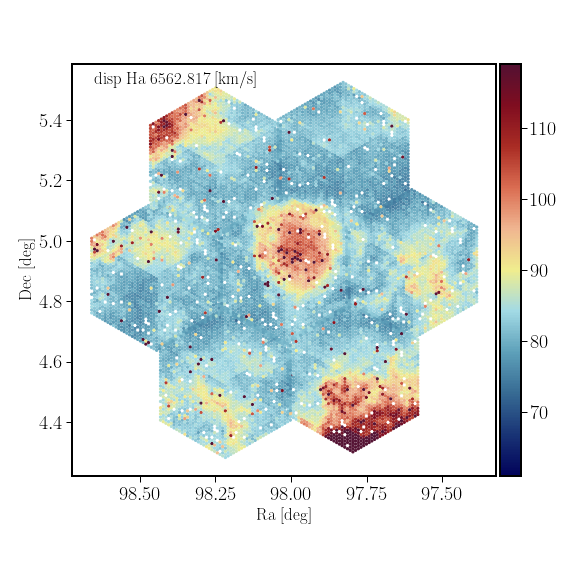}\includegraphics[width=6cm,trim={10 55 20 50},clip]{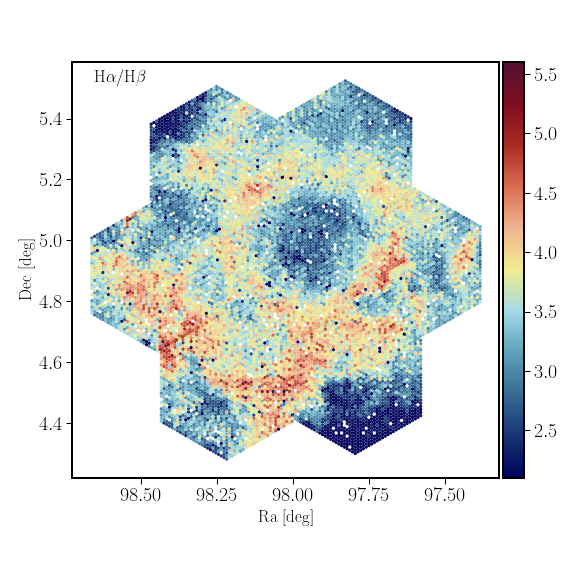}
 \includegraphics[width=6cm,trim={10 55 20 50},clip]{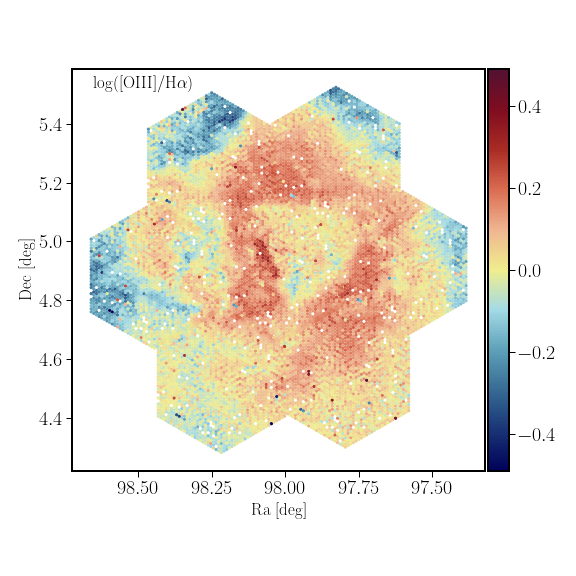}\includegraphics[width=6cm,trim={10 55 20 50},clip]{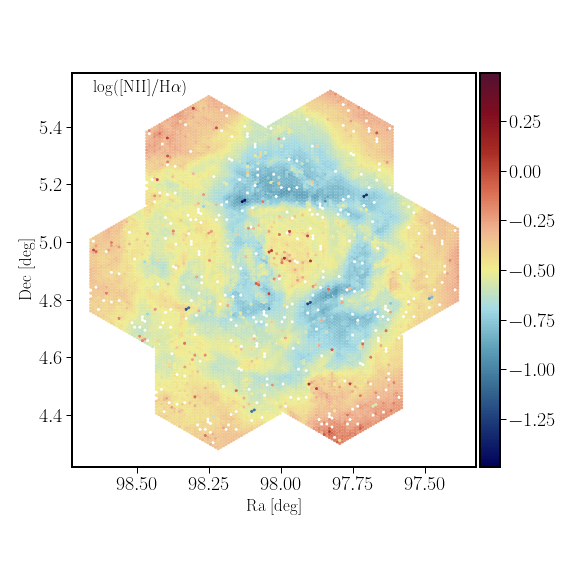}\includegraphics[width=6cm,trim={10 55 20 50},clip]{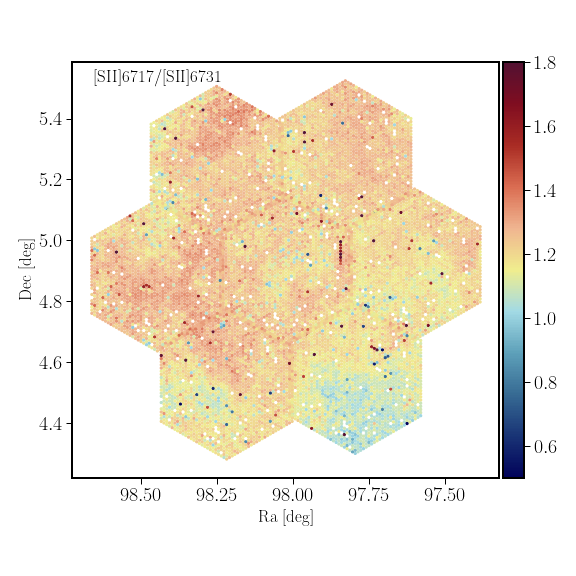}
 \endminipage
\caption{Spatial distribution of different parameters derived from the weighted-moments analysis included in the LVM pDAP for the Rosette nebulae data shown in Fig.~\ref{fig:dap_RT1}. From top-left to bottom-right we show (i) the H$\alpha$ velocity, (ii) the H$\alpha$ velocity dispersion, (iii) the H$\alpha$ to H$\beta$ flux intensity ratio, (iv) the [\ion{O}{III}] to H$\beta$ flux intensity ratio, (v) the [\ion{N}{II}] to H$\alpha$ flux intensity ratio and (vi) the [\ion{S}{II}]6717 to [\ion{S}{II}]6731 flux intensity ratio.}\label{fig:dap_RT2}
\end{figure*}


Figure~\ref{fig:dap_spec} illustrates the pDAP analysis using data in the Rossette Nebula. The best-fit final model generated by the pDAP is shown in the upper panel. The middle panels show a set of zoom-ins highlighting the fits to the strongest emission lines using multi-Gaussian models, including [\ion{O}{II}]3726,28 , H$\delta$, H$\gamma$, H$\beta$, [\ion{O}{III}]5007, H$\alpha$+[\ion{N}{II}]6548,83, [\ion{S}{II}]3717,31, [\ion{S}{III}]9069 and [\ion{S}{III}]9531. These emission lines are frequently used to determine the properties of the ionized gas, including the ionizing source, the dust attenuation, the electron density, the ionization parameter and strength of the ionization \citep[e.g.][]{Vilchez1988,kewley01}. The bottom row comprises two panels illustrating the quality of the stellar template fit (3800-4100\AA\ and 4800-5100\AA), and four additional ones showing the detection of weak but relevant emission lines: (i) HeI4471 and HeI6678, tracers of the He abundance \citep[e.g., ][]{valerdi2021}, (ii) [\ion{O}{II}]7318, a tracer of super-nova remanants \citep[SNR, e.g.,][]{CF2021}, and (iii) [\ion{N}{II}]5755 and [\ion{S}{III}]6312 \citep[e.g.,][]{marino13}, sensitive to the electron temperature. There are computational and accuracy reasons why not all the emission lines are fit using parametric models, extensively discussed in previous studies \citep[e.g.,][]{pipe3d_ii,pypipe3d,sanchez22}. A typical example of such lines are auroral lines, a particularly interesting set of emission lines for the science goals of the LVM, as they are frequently used to trace the electron temperature of different species (e.g., [\ion{N}{II}]5755 and [\ion{S}{III}]6312, shown in bottom panels in Fig. \ref{fig:dap_spec}). For these weak lines, the weighted-moment analysis provides reliable estimations of their properties instead. The wavelengths of the full set of emission lines analyzed are shown as vertical orange lines in all panels of Fig.~\ref{fig:dap_spec}. We are currently exploring the use of both Gaussian fits and non-parametric estimates in our analysis pipeline. 

Figures~\ref{fig:dap_RT1} and \ref{fig:dap_RT2} showcase spatially-resolved properties of the ionized gas in the LVM data. Each panel shows the spatial distribution of a particular parameter extracted from a seven-pointing mosaic of the Rosette nebula, 12,607 spectra in total, obtained during early-science observations in October 2023. Fig.~\ref{fig:dap_RT1} shows the flux intensity maps for a sub-set of analyzed emission lines, including the [\ion{O}{II}]3728, H$\gamma$, H$\beta$, [\ion{O}{III}]5007, H$\alpha$, [\ion{N}{II}]6583, [\ion{S}{II}]6717, [\ion{S}{II}]6731, and  [\ion{S}{III}]9531. The changes in the ionizing conditions and physical properties of the nebula are clearly seen as different spatial distributions. It is important to highlight that despite the considerable dynamic range of the fluxes, we detect emission in all the emission lines in this figure at any location within the nebula.

Fig.~\ref{fig:dap_RT2} shows the spatial distribution of some additional properties of the ionized gas emission lines. The first two panels, from top-left to bottom right, shows the velocity and velocity dispersion derived for H$\alpha$. Both figures highlights the accuracy achieved by the LVM data to measure low-velocity patterns. We are measuring velocities with an accuracy of a fraction of 1\,km\,s$^{-1}$, and velocity dispersions with an accuracy of the order of a few km\,s$^{-1}$. The remaining panels of this figure show the spatial distribution of a set of line ratios frequently utilized in the analysis of ionized gas. The third panel shows the H$\alpha$ to H$\beta$ ratio, a tracer of the dust attenuation \citep[e.g.,][]{osterbrock89, wild11,salim20}. A ring structure and a shell to the south-east are clearly observed. The fourth and fifth panels show the [\ion{O}{III}]/H$\beta$ and [\ion{N}{II}]/H$\alpha$ line ratios, frequently used to trace the ionizing source (for unresolved nebulae) and the changes of the ionizing conditions (for resolved ones). There is a clear and expected anti-correlation between these parameters, with regions of high [\ion{O}{III}]/H$\beta$ corresponding to values of low [\ion{N}{II}]/H$\alpha$. Finally, the sixth panel shows the [\ion{S}{II}]6716/[\ion{S}{II}]6731 line ratio, a tracer of the varying electron density \citep[e.g.,][]{osterbrock89}.


\subsection{Data Products and Releases}

As has been the case for two decades, all SDSS-V data will be made public. LVM is no exception, and the first public data release of LVM raw, reduced, and science-ready data and products will happen with SDSS DR20 in late 2025 and yearly after that, with the full dataset becoming public one year after survey operations end.

Among the data we will release are the raw data; fully reduced, sky subtracted, and flux calibrated row-stacked spectra (RSS files) including fluxes, uncertainties, LSF information, PSF information, fiber RA and DEC, and sky spectra; fluxes of more than 200 emission lines or upper limits; kinematic data for subsets of those emission lines; as well as all code used to reduce and analyse the data, map and cube making routines; and finally all template spectra used in reduction and analysis. Model-dependent quantities such as densities, temperatures, abundances, fits to nebular emission models, etc., will be made available as Value Added Catalogs (VACs).

\section{Summary} \label{sec:summary}

We present the Local Volume Mapper survey, part of the SDSS-V project. The LVM is a new observing facility, instrument, and survey project covering $>4300$\,square degrees of sky with IFU observations, many orders of magnitude larger than any previous effort. 
The LVM aims to map the ionized ISM of the Milky Way and the Magellanic Clouds on global scales of galactic outflows, fountains and winds, down to individual sources of feedback. LVM is resolving the scales on which energy, momentum, and metals are injected into the ISM (mainly from high-mass stars) over scales of gas clouds, and scales where energy is being dissipated (cooling, shocks, turbulence, bulk flows) to global scales that couple to Galactic kinematics, disk-scale structures such as the bar and spiral arms, and finally gas in and outflows.

To achieve these goals, the LVM will carry out an integral-field spectroscopic survey of the Galaxy, LMC, SMC, and a sample of local volume galaxies utilizing a new survey facility consisting of alt-alt mounted celiostats feeding 16\,cm aperture refractive telescopes, a lenslet-coupled fiber-optic system and an array of spectrographs, covering 3600-9800\,\AA\ at $R\sim4000$. The ultra-wide field science IFU has a diameter of 0.25\,degrees with 1801 35.5\arcsec clear apertures in hexagonal arrangement of 25 rings. The celiostat concept allows for stationary powered optics and most importantly a completely stationary fiber system, avoiding many problems in stability of the line spread function seen in traditional fiber systems \citep{law21,bundy22}. This innovative instrument will conduct a survey delivering more than $55\times10^6$ spectra spanning the bulk of the Milky Way disk visible from Las Campanas Observatory at spatial resolutions of 0.05 to 1\,pc (depending on distance), the Magellanic Clouds at 10\,pc resolution with over $10^6$ resolution elements on the LMC, and a sample of very nearby galaxies of large apparent diameter. 

All observations in the MW have a nominal exposure time of 900\,s, to reach a line sensitivity of $5\sigma=6\times10^{-18}$\,erg\,s$^{-1}$\,cm$^{-2}$\,arcsec$^{-2}$ at 6563\,\AA. The Milky Way is observed at any lunation, with observing constraints of distance to the moon larger than 60$\degr$, airmass below 1.75, and shadow height of $>1000$\,km to lower the impact of geocoronal \Halpha\ on the MW signal. All observations in the Magellanic System have a nominal exposure time of 8100\,s broken up into 9 exposures of 900\,s each. The exposures are dithered on a 9 point grid similar to the one employed in MaNGA (see \citealp{law16}).
The Magellanic System observations reach an emission line depth of $5\sigma=2\times10^{-18}$\,erg\,s$^{-1}$\,cm$^{-2}$\,arcsec$^{-2}$ at 6563\,\AA, and a 
continuum surface brightness depth at $5\sigma$ of 23.0\,AB\,mag\,arcsec$^{-2}$ in the V band. The Magellanic System surveys are covering the LMC and SMC roughly to the galaxies' optical radius, $D_{25}$.

LVM has started survey operations in November 2023 and is planning on observing through May 2027. As was the case with all previous and current surveys in SDSS, all raw and reduced data as well as code will be made public, starting with SDSS DR20 in late 2025. Early science data have already shown that the LVM is able to deliver the  high quality spectral mapping of Galactic nebulae necessary to address our core science goals and provide a new view of the energy injection scale within Local Group galaxies.

\begin{acknowledgements}
Funding for the Sloan Digital Sky Survey V has been provided by the Alfred P. Sloan Foundation, the Heising-Simons Foundation, the National Science Foundation, and the Participating Institutions. SDSS acknowledges support and resources from the Center for High-Performance Computing at the University of Utah. SDSS telescopes are located at Apache Point Observatory, funded by the Astrophysical Research Consortium and operated by New Mexico State University, and at Las Campanas Observatory, operated by the Carnegie Institution for Science. The SDSS web site is \url{www.sdss.org}.

SDSS is managed by the Astrophysical Research Consortium for the Participating Institutions of the SDSS Collaboration, including Caltech, The Carnegie Institution for Science, Chilean National Time Allocation Committee (CNTAC) ratified researchers, The Flatiron Institute, the Gotham Participation Group, Harvard University, Heidelberg University, The Johns Hopkins University, L’Ecole polytechnique f\'{e}d\'{e}rale de Lausanne (EPFL), Leibniz-Institut f\"{u}r Astrophysik Potsdam (AIP), Max-Planck-Institut f\"{u}r Astronomie (MPIA Heidelberg), Max-Planck-Institut f\"{u}r Extraterrestrische Physik (MPE), Nanjing University, National Astronomical Observatories of China (NAOC), New Mexico State University, The Ohio State University, Pennsylvania State University, Smithsonian Astrophysical Observatory, Space Telescope Science Institute (STScI), the Stellar Astrophysics Participation Group, Universidad Nacional Aut\'{o}noma de M\'{e}xico, University of Arizona, University of Colorado Boulder, University of Illinois at Urbana-Champaign, University of Toronto, University of Utah, University of Virginia, Yale University, and Yunnan University.

KK, OE, EE, JL, NS and JEMD gratefully acknowledge funding from the Deutsche Forschungsgemeinschaft (DFG, German Research Foundation) in the form of an Emmy Noether Research Group (grant number KR4598/2-1, PI Kreckel) and the European Research Council’s starting grant ERC StG-101077573 (“ISM-METALS"). 
K.H.\ acknowledges support from ANID -- Millennium Science Initiative Program -- Center Code NCN2021\_080.
RSK and SCOG acknowledge funding from the ERC via Synergy Grant ``ECOGAL'' (project ID 855130), from the German Excellence Strategy via the Heidelberg Cluster of Excellence (EXC 2181 - 390900948) ``STRUCTURES'', and from the German Ministry for Economic Affairs and Climate Action in project ``MAINN'' (funding ID 50OO2206). They also acknowledge computing resources provided by the State of Baden-W\"{u}rttemberg and DFG through grant INST 35/1134-1 FUGG and data storage at SDS@hd through grant INST 35/1314-1 FUGG. G.A.B., EJJ and B.D. acknowledge the support from the ANID Basal project FB210003.   EJJ acknowledges financial support of Millenium Nucleus ERIS NCN2021\_017 and ANID-FONDECYT iniciación grant No.\ 11200263. B.D. acknowledges support by ANID-FONDECYT iniciación grant No.\ 11221366.

\end{acknowledgements}

\bibliography{lvm_overview}{}
\bibliographystyle{aasjournal}



\end{document}